\documentclass{aa}

\usepackage{amsmath}
\usepackage[utf8]{inputenc}
\usepackage{natbib}
\usepackage{graphicx}
\usepackage{txfonts}
\usepackage[draft]{hyperref}
\usepackage{color}
\usepackage{upquote} 
\usepackage[normalem]{ulem}
\usepackage{xspace}
\usepackage{xcolor}

\def\Gaia{\textit{Gaia}\xspace}
\def\gmag{\ensuremath{G}\xspace}
\def\gbp{\ensuremath{G_{\rm BP}}\xspace}
\def\grp{\ensuremath{G_{\rm RP}}\xspace}

\newcommand{\ks}{\ensuremath{K_{\rm s}}\xspace}
\newcommand{\wjk}{\ensuremath{W_{\rm J,K_s}}\xspace}
\newcommand{\wrp}{\ensuremath{W_{\rm BP,RP}}\xspace}
\newcommand{\Msun}{\ensuremath{{\rm M}_{\odot}}\xspace}
\newcommand{\mbol}{\ensuremath{m_{\rm bol}}\xspace}

\begin{document}

\title{Period-luminosity diagram of long period variables in the Magellanic Clouds}
\subtitle{New aspects revealed from \Gaia Data Release 2.}
  \titlerunning{Period-luminosity diagram of LPVs in \Gaia DR2}
  \authorrunning{Lebzelter et al.}

\author{T. Lebzelter\inst{\ref{inst_vie}}\fnmsep\thanks{Corresponding author: T. Lebzelter
(\href{mailto:thomas.lebzelter@univie.ac.at}{\tt thomas.lebzelter@univie.ac.at})},
M.~Trabucchi\inst{\ref{inst_pdv}},
N.~Mowlavi\inst{\ref{inst_gen}},
P.R.~Wood\inst{\ref{inst_mso}}, P.~Marigo\inst{\ref{inst_pdv}},
G.~Pastorelli\inst{\ref{inst_pdv}},
I.~Lecoeur-Ta\"{i}bi\inst{\ref{inst_gen}}
}
 \institute{University of Vienna, Department of Astrophysics, Tuerkenschanz\-strasse 17, A1180 Vienna, Austria\label{inst_vie}
 \and 
 Dipartimento di Fisica e Astronomia Galileo Galilei, Università di Padova, Vicolo dell’Osservatorio 3, I-35122 Padova, Italy\label{inst_pdv}
 \and 
 Department of Astronomy, University of Geneva, Ch. des Maillettes 51, CH-1290 Versoix, Switzerland\label{inst_gen}
 \and 
 Research School of Astronomy and Astrophysics, Australian National University, Canberra, ACT 2611, Australia\label{inst_mso}
       }
\date{July 2019}

\abstract{
          The period-luminosity diagram (PLD) has proven to be a powerful tool for studying populations of pulsating red giants.
          \Gaia Data Release 2 (DR2) provides a large data set including many long-period variables (LPVs) on which this tool can be applied. 
         }
         {
          We investigate the location of LPVs from the Large and Small Magellanic Clouds in the PLD using various optical and infrared luminosity indicators from \Gaia and 2MASS, respectively.
          We thereby distinguish between stars of different masses and surface chemistry.
         }
         {
          The data set taken from the \Gaia DR2 catalogue of LPVs allows for a homogeneous study from low- to high-mass LPVs.
          These sources are divided into sub-populations of asymptotic giant branch (AGB) stars according to their mass and their O- or C-rich nature using the \Gaia-2MASS diagram developed by our group.
          This diagram uses a Wesenheit index \wrp based on Wesenheit functions in the \Gaia and 2MASS photometric bands.
          Four different luminosity indicators are used to study the period-luminosity (P-L) relations.
         }
         {
          We provide the first observational evidence of a P-L relation offset for both fundamental and 1O pulsators between low- and intermediate-mass O-rich stars, in agreement with published pulsation predictions.
          Among the luminosity indicators explored, sequence C' is the narrowest in the P-\wrp diagram, and is thus to be preferred over the other PLDs for the determination of distances using LPVs.
          The majority of massive asymptotic giant branch (AGB) stars and red supergiants form a smooth extension of sequence C of low- and intermediate-mass AGB stars in the P-\wrp diagram, suggesting that they pulsate in the fundamental mode.
          All results are similar in the two Magellanic Clouds.
         }
         {
         }

\maketitle

\section{Introduction}
Long period variables (LPVs) are easily detectable representatives of the asymptotic giant branch (AGB) phase of the evolution of low- and intermediate mass stars.
The AGB phase is critical for stellar and galactic evolution \citep[e.g.][]{pastorelli_etal_2019, marigo_etal_2017, marigo_15, Bruzual_07, Maraston_etal_06}.
In terms of stellar evolution, high mass-loss rates are responsible for the end of the life of a star.
At the same time, a large fraction of the intermediate to heavy elements are produced and returned to the interstellar medium during the AGB phase \citep[e.g.][]{Ventura_etal_18, Slemer_etal_17, Cristallo_etal_15, karakas_lattanzio_2014}.
Therefore, this phase is of high importance for the chemical evolution of galaxies.

Long period variables are characterised by large amplitude variations in the visual and pulsation periods of about 10 to 1000 days. 
The period-luminosity diagram (PLD) has become one of the main tools for studying LPVs and their evolution along the AGB. 
\citet{wood_sebo_1996} were the first to detect that semi-regular variables (SRVs) follow an additional P-L relation other than that of Miras. This finding was interpreted, and later confirmed,  as due to pulsations in overtone modes, while Miras pulsate in the fundamental mode \citep[e.g.][]{wood_etal_1999,wood_2000}.
As a consequence of the lack of reliable distances to LPVs in the solar neighbourhood, studies of the PLD have relied on variables in stellar systems, most of all the Magellanic Clouds \citep{feast_et_al_1989, wood_2000, ita_et_al_2004a, fraser_et_al_2005, soszynski_etal_2007}, but also in globular clusters \citep{feast_et_al_2002, lebzelter_wood_2005}, the Galactic 
Bulge \citep{wood_bessell_1983, glass_et_al_1995} and some local
group galaxies \citep{whitelock_2012, whitelock_et_al_2013, menzies_et_al_2015}.

Studies of the Magellanic Clouds finally revealed the presence of at least five parallel P-L relations, or sequences \citep{wood_etal_1999,ita_etal_2004}. 
Distinct sequences are associated with distinct pulsation modes, with two exceptions: sequences B and C$^{\prime}$ are both due to the same pulsation mode \citep[the first overtone (1O) mode;][]{trabucchi_etal_2017}, while sequence D hosts the so-called long secondary periods (LSPs), which are likely the result of a different kind of variability, but whose origin is still unknown \citep{hinkle_etal_2002, wood_nicholls_2009, saio_etal_2015}. 
\citet{trabucchi_etal_2017} argued that when stars pulsating predominantly in the 1O mode are crossing the region between sequences B and C$^{\prime}$, the stars tend to develop a LSP with a larger amplitude than the 1O mode itself. 
The LSP is thus more easily detected, resulting in the apparent gap between sequences B and $C^{\prime}$.
Yet, despite pulsating in the same mode, stars associated with sequence $C^{\prime}$ (generally classified as SRVs) have observed properties markedly distinct from those on sequence B \citep[the so-called OGLE\footnote{Optical Gravitational Lensing Experiment} Small Amplitude Red Giants, OSARGs;][]{wray_etal_2004}.
The SRVs not only show larger visual amplitudes, but they also have higher infrared excess and estimated mass-loss rates \citep{mcdonald_trabucchi_2019}. 
In this scenario, Mira variables are found to lie on sequence C, associated with pulsation in the fundamental mode and are experiencing the latest stages of the AGB.

\begin{figure*}
\centering
\includegraphics[width=\textwidth]{{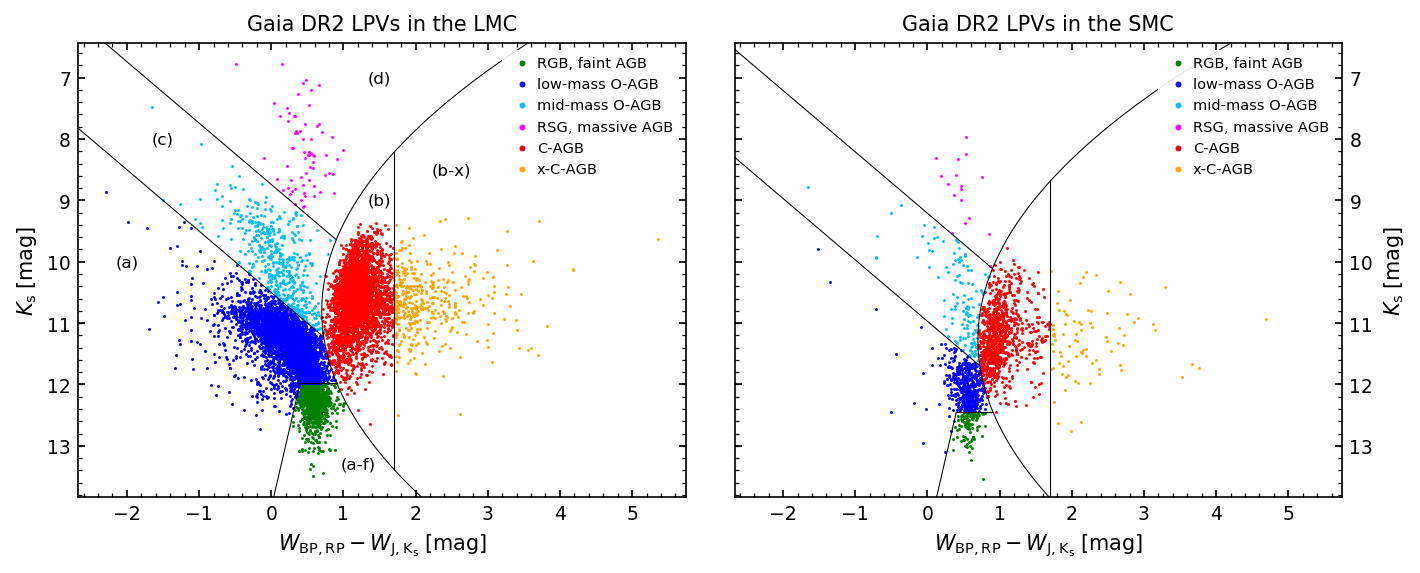}}
\caption{\Gaia-2MASS diagram of LPVs in the LMC (left) and SMC (right). Boundaries between various groups of stars are defined as in \citet{lebzelter_etal_2018}, except for the redefined boundary between branches (c) and (d) (see Sect.\,\ref{samplechar}).
Boundaries of the SMC account for the different distance with adopted distance moduli $18.49$ mag and $18.96$ mag, respectively \citep{degrijs_etal_2017}.
        }
\label{fig:LMC_SMC_G2MD_apparent}
\end{figure*}
 
The past 20 years of research in this area revealed the dominant role of large surveys for the detection and study of P-L relations. 
Among the ground-based studies, the monitoring of the Magellanic Clouds and the galactic bulge within the OGLE project \citep{soszynski_etal_2009_ogle3lmc,soszynski_etal_2011_ogle3smc,soszynski_etal_2013_ogle3bulge} stands out because of the long time coverage of light changes, the high time resolution and the sky area it comprises. 
With the advent of \Gaia Data Release 2 (DR2), a database became available which provides a deep all-sky monitoring of LPVs in one wide (\gmag) and two semi-wide (\gbp and \grp) optical bands covering bright and faint stars from few mags down to $\gmag=20.7$~mag, and with distances and proper motions as additional information.
The 22 months of data published in DR2 already allow for a study of periods and pulsational behaviour of LPVs \citep{mowlavi_etal_2018_dr2lpv}.

Recently, our group showed \citep{lebzelter_etal_2018} that a combination of Wesenheit functions using \Gaia and near-infrared wavelength bands allows us to efficiently separate groups of stars on the AGB according to their mass and surface chemistry (C/O-rich). 
To simplify the description, in the following, a diagram plotting 2MASS $\ks$ against the above-mentioned difference of Wesenheit indices (see Sect.\,\ref{samplechar} for the definitions of $W_{BP,RP}$ and $W_{J,K}$) is called a \Gaia-2MASS diagram.
This diagram allows the possibility to add a mass and dredge-up indicator to the PLD.
A further exploration of the applicability of this diagram to various stellar systems has been presented in \citet{eslab53_Mowlavi_2019}.

Pulsation models of LPVs predict a dependency of the P-L relations on mass \citep{trabucchi_etal_2019} and surface chemistry \citep[e.g.][]{lebzelter_wood_2007}.
The goal of this paper is to extend our understanding of variability on the AGB, particularly for the more massive and thus more luminous part of this evolutionary phase, and in relation to the distinction between C- and O-rich atmospheric chemistry.
We investigate PLDs using \Gaia photometry as brightness indicators to add these relations to the suite of tools for the exploration of LPVs.
At the same time, we aim to explore further the strengths of the \Gaia-2MASS diagram for the study of evolved stellar populations.


\section{Sample}
\subsection{Selection and characterization of sample stars}\label{samplechar} 

\begin{figure*}
\centering
\includegraphics[width=\textwidth]{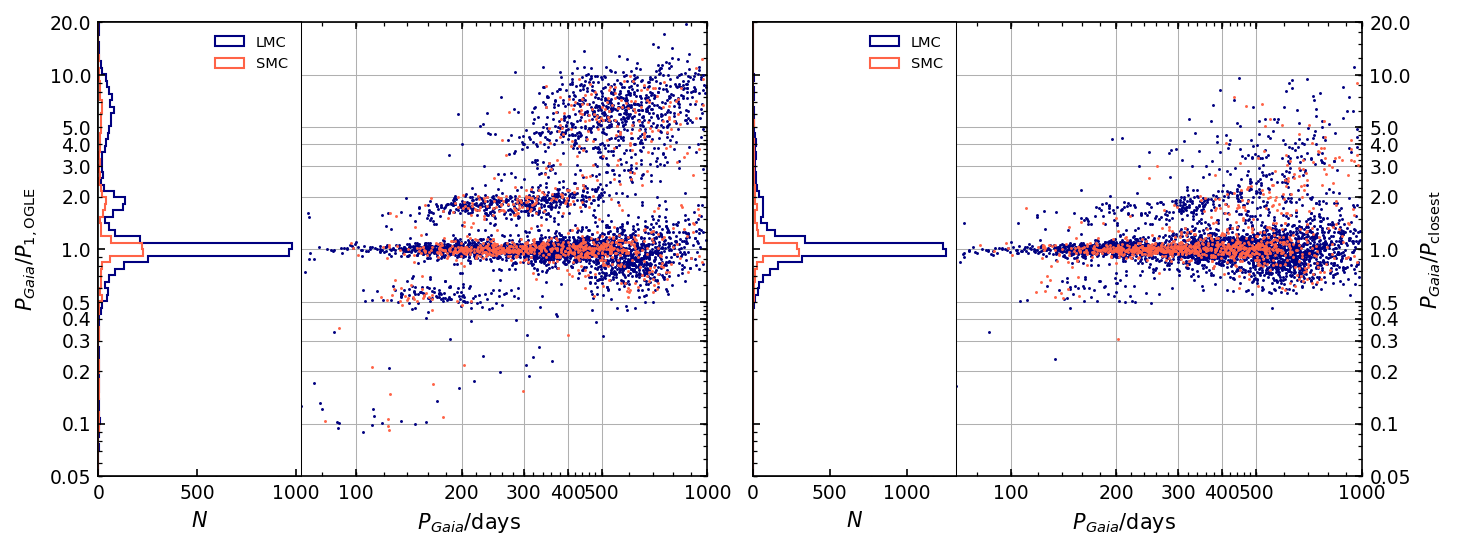}
\caption{Left panel: Ratio of the \Gaia DR2 period to the primary period from the OGLE-III database for the LMC (blue) and SMC (red), respectively, as well as the corresponding histogram. Right panel: Same, but using the OGLE-III period whose value is closest to the \Gaia DR2 period. Above $P_{Gaia}=500$ days, \Gaia periods become uncertain owing to the limited time span of the \Gaia DR2 data.
        }
\label{Fig:MCs_Gaia_OGLE_periods}
\end{figure*}

The starting point for our sample selection was the \Gaia DR2 database of LPVs \citep{mowlavi_etal_2018_dr2lpv} because for these objects the availability of \gbp and \grp photometry allows for a distinction according to mass using the \Gaia-2MASS diagram and because \Gaia also covers the brightest objects in the Magellanic Clouds, which OGLE does not.
We selected stars in the LMC and SMC similar to the selection process in \citet{lebzelter_etal_2018} and \citet{eslab53_Mowlavi_2019} according to their location in the sky, their proper motion, and their parallax.
While we present results for both clouds in this paper, the description and interpretation focusses on the LMC owing to the much lower number of sample stars in the SMC.

The sample inherits the limitations of the first \Gaia catalogue of LPVs, which focussed on candidates with variability amplitudes larger than 0.2~mag in \gmag, and prioritised low contamination over completeness, resulting in a catalogue of LPV candidates with a completeness of about 45\% in the Magellanic Clouds.
In addition, the study of the periods in this paper is restricted to the range between 70 and 1000 days, the lower limit resulting from restrictions due to the \Gaia scanning law introducing spurious frequencies in the low end, and the upper limit resulting from the 22-month time span of \Gaia DR2 \citep[we refer to][for a description of the properties of the catalogue]{mowlavi_etal_2018_dr2lpv}.
While the latter condition would in principle limit the period range to a maximum of $\sim$500~d, we still extend the study up to 1000~d to include LPVs that show variations on that timescale, particularly LSPs forming sequence D in the PLD, keeping in mind that the uncertainties on the periods longer than 500~d can be large.
Finally, we note that the DR2 catalogue of LPVs includes only the variability period of the largest amplitude. Multi-periodicity, while being very common among LPVs, is awaiting additional data in future data releases to be properly characterised.

These limitations, in combination with a higher limiting amplitude, substantially affect the appearance of the PLD, especially when compared with OGLE observations. The lower period limit, in particular, effectively leads to the loss of the low-brightness parts of the P-L sequences C$^{\prime}$ and C (cf. Sect.~\ref{sec:results}). In addition, stars with a 1O mode period transitioning between sequences B and C$^{\prime}$ are likely to develop a more easily detectable LSP on sequence D. When, as in the present case, a single observed period per star is available, this contributes to the apparent depletion of sequence C$^{\prime}$.

We then attributed classifiers for mass (low, intermediate, and high) and chemical type (O-rich or C-rich) to each of our LPVs using the \Gaia-2MASS diagram and the definitions for the LMC set up in \citet{lebzelter_etal_2018}.
This is illustrated in the left panel of Fig.~\ref{fig:LMC_SMC_G2MD_apparent}, in which the magnitude in the 2MASS $\ks$ band is shown against the difference between the Wesenheit indices constructed with \Gaia and 2MASS photometry, respectively,
i.e.\begin{equation}\label{Wrpdef}
    \wrp = \grp - 1.3 \left( \gbp - \grp \right);
\end{equation}
\begin{equation}\label{WJKdef}
    \wjk = \ks - 0.686 \left( J - \ks \right).
\end{equation}

\citet{lebzelter_etal_2018} identified four branches in the \Gaia-2MASS diagram, labelled with letters from (a) to (d). 
In this work, we use the same labels as in \citet{lebzelter_etal_2018} to identify the four branches in the \Gaia-2MASS diagram. 
In addition, we use the label (a-f) to identify the portion of branch (a) populated by RGBs and faint AGBs (i.e. stars below the tip of the RGB), and label (b-x) for the right-most side of branch (b), containing stars identified as 
extremely dusty stars. 
That group predominantly contains C-stars, which are efficiently producing high opacity dust grains \citep{hoefner_olofsson_2018}, but may also include highly reddened OH-IR stars.
Using the list of OH-IR stars of the LMC recently presented by \citet{goldman_et_al_2017} we find nine OH-IR stars among the oxygen rich stars in our diagram, and six objects in the part denoted as (b-x).
All but one of the latter OH-IR stars are among the reddest objects in (b-x).

We applied one modification to the definitions of the various groups in this diagram compared to \citet{lebzelter_etal_2018}, namely for the boundaries for groups (c) (intermediate-mass) and (d) (high-mass), in which case we replaced the original definitions with, respectively,
\begin{equation}
    K_{\rm s} > 8.74 + (\wrp - \wjk)\;{\rm and}\;K_{\rm s}\leq10.5+(\wrp-\wjk)\,;
\end{equation}
\begin{equation}
    K_{\rm s}\leq 8.74 + (\wrp - \wjk)\,.
\end{equation}
This new boundary between groups (c) and (d) accounts better for the distribution of LPVs in the SMC (right panel of Fig.~\ref{fig:LMC_SMC_G2MD_apparent}) while keeping a good solution for the LMC.
A possible dependency of the boundaries on metallicity still needs to be explored.

The \Gaia-2MASS diagram of the SMC shown in Fig.~\ref{fig:LMC_SMC_G2MD_apparent} accounts for the difference in distance with the LMC assuming a distance modulus $18.49\pm0.09$~mag for the LMC and $18.96\pm0.02$ ~mag for the SMC \citep{degrijs_etal_2017}.
It is clear that the chosen values of the distance moduli affect the classification of stars that are located nearby boundaries between groups, thereby introducing a source of error. 
This uncertainty overlaps with instrumental uncertainties, which are in the millimagnitude range for the \Gaia photometry \citep{evans_etal_2018} and about 0.03\,mag for the 2MASS data.
However, because of the limited sampling of the light curves, deviation from the mean brightness of a star, owing to its variability and cycle-to-cycle fluctuations,  certainly dominates the uncertainty in the two quantities plotted in Fig.~\ref{fig:LMC_SMC_G2MD_apparent}.
Naturally, this leads to misclassified objects.
Additional sources of error stem from the different extinction between the two systems (here neglected) as well as the different magnitude of the RGB tip, whose value was employed by \citet{lebzelter_etal_2018} to discriminate between low-mass O-rich AGBs and the group consisting of RGBs and faint AGBs. 
We do not consider this a problem in the present paper since we do not aim to derive statistical quantities from our analysis here.

The bolometric corrections (BCs) for each object in our sample are solely based on \Gaia data (\gmag, \gbp, \grp).
These corrections differ from those used in DR2 and are described in detail in Appendix~\ref{sec:bc}.
These BCs are used to compute the bolometric magnitudes \mbol from the \gmag-band photometry.

\subsection{Comparing OGLE and \Gaia periods}
As mentioned above, the OGLE survey provides one of the best catalogues of LPVs in the Magellanic Clouds in terms of completeness and accuracy and
listing multiple periods and amplitudes.
Accordingly, OGLE data have been widely used to study PLDs of LPVs, in particular for the study of \citet{trabucchi_etal_2017}. 
For our study of the pulsational properties of LPVs with \Gaia data, it seemed self-evident to use OGLE data as a reference to judge the accuracy, in particular, of the \Gaia periods and to compare any conclusions deduced from our study to previous works on the PLD. 
A first test of that kind was part of the paper by \citet{mowlavi_etal_2018_dr2lpv} indicating a reasonable reproduction of OGLE periods and amplitudes with the help of \Gaia data.
We note that there are two advantages of using \Gaia data for our study. First, it is possible to access the stellar parameters mass and chemistry (using the \Gaia-2MASS diagram) and, second, the data include bright AGB stars in the Magellanic Clouds, which have been cut off due to the upper brightness limit of the OGLE survey. The disadvantages of \Gaia
data are the shorter lengths of the time series, lower sampling density of the light curves, and -- for DR2 data -- the availability of only one single period.

\begin{figure}
    \centering
    \includegraphics[width=\columnwidth]{{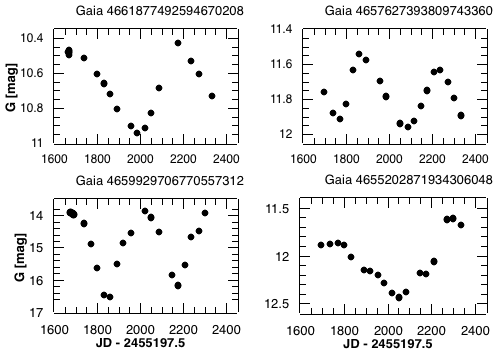}}
    \caption{Example \Gaia light curves of four group [d] stars. Top row: Two representative supergiants. Bottom left: Large amplitude variable, likely a massive AGB star. Bottom right: Incomplete light curve coverage to determine a period reliably.}
    \label{fig:rsg_example}
\end{figure}

\begin{figure*}
\centering
\includegraphics[width=\textwidth]{{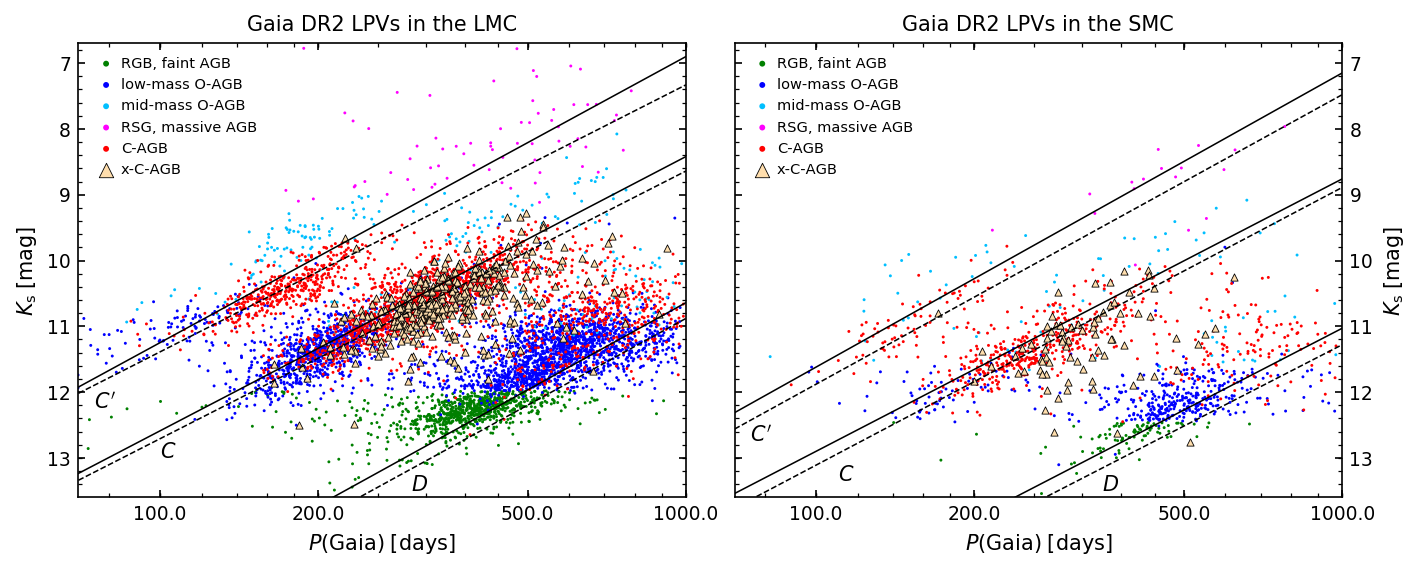}}
\caption{Period-luminosity diagram in the 2MASS $\ks$ band of LPVs in the LMC (left) and SMC (right). Stars are colour-coded according to the group they belong to as in Fig.~\ref{fig:LMC_SMC_G2MD_apparent}. Best-fit relations to O-rich (solid lines) and C-rich (dashed lines) are taken from \citet{soszynski_etal_2007}.
        }
\label{fig:PLD_logP_Ks_LMC_SMC}
\end{figure*}

We compare periods between the two surveys in Fig.~\ref{Fig:MCs_Gaia_OGLE_periods}. First, we compare \Gaia periods with the primary periods (i.e. those with the largest amplitude in a given star) detected by OGLE\footnote{
    We cross-matched the OGLE and \Gaia DR2 catalogues to 1 arcsec of radius and found a match for 5279 of the total 12811 Magellanic Cloud LPVs from the \Gaia DR2 data set (the majority of the remaining LPVs are out of the OGLE field of view).
}. 
The ratio between the two is plotted against the \Gaia period in the left panel of Fig.~\ref{Fig:MCs_Gaia_OGLE_periods}, which also shows the histogram of these ratios. 
We find a high rate of agreement; about 70\% of the \Gaia DR2 periods in the range 70-500 days, differing by less than 25\% from OGLE primary periods. 
At longer periods, the limited length of the \Gaia time series introduces a larger period uncertainty. 
For a significant fraction of stars, there are evident discrepancies between the two surveys, a direct consequence of multi-periodicity and of the lower period limit of the \Gaia catalogue. 
Indeed,  a multi-periodic star with a primary period shorter than 60 days is present in the OGLE catalogue, but not in \Gaia , which lists one of the  longer periods of the star instead. 
This is the case for objects in the upper right corner in the left panel of Fig.~\ref{Fig:MCs_Gaia_OGLE_periods}, for which the \Gaia pipeline detected a LSP in place of, for example the 1O mode period found by OGLE. 
Similarly, the 1O mode period may be listed in the present catalogue in place of the fundamental mode period or vice versa.
Figure 2 shows candidates with this difference to the OGLE results in the two groups located directly above and below  P$_{Gaia}$/P$_{OGLE}$=1. 
A large fraction of objects disappear from these two groups when switching to P$_{Gaia}$/P$_{closest}$, i.e. when using the OGLE period nearest to the \Gaia period (right panel of Fig.~\ref{Fig:MCs_Gaia_OGLE_periods}).
In the remaining cases the \Gaia period search algorithm seems to have picked the first harmonic of the OGLE period, likely because of the sampling of the light curve by \Gaia which is very different from OGLE. 
The bias introduced by multi-periodicity can thus be partially accounted for by comparing \Gaia periods with the closest period listed for the same star in the OGLE catalogue, rather than the primary OGLE period. 

Our comparison proves that within the investigated period range, \Gaia periods are in very good agreement with the OGLE periods. Therefore, the use of \Gaia light curve data in this study is a valid approach to investigate the PLDs of LPVs. Since we restrict our conclusions on the variability found on sequence D to qualitative statements, accuracy limitations of \Gaia data near the long period end play less of a role. The same is true for the bias associated with the lack of multi-periodicity information in the \Gaia catalogue.

The high-mass stars and supergiants [group (d)] require a further discussion to verify the quality of the periods for this group. 
The red supergiants, as the most luminous objects in the PLD, naturally show, on average, the longest periods and populate a period range where the length of the \Gaia DR2 data set is of the same order as the period. 
We show in Fig.~\ref{Fig:MCs_Gaia_OGLE_periods}  that periods above 500\,days derived from the \Gaia DR2 data exhibit an increasing uncertainty.
However, this finding is based on a general comparison of \Gaia and OGLE data.
The OGLE observations are saturated for the supergiants in the Magellanic Clouds and therefore a direct comparison with our group (d) stars is not possible except for those with the lowest brightnesses.
Hoping to find an alternative sample for comparison, we extracted light curve data for our supergiants from the Massive Compact Halo Objects Project (MACHO) database.
However, it turned out that those measurements seem to be affected by saturation effects as well and do not provide conclusive results. 
Owing to the lack of usable comparison data we decided to look at the individual light curves to evaluate the quality of the derived periods.
This exercise confirmed the variable nature of our sample supergiants.
Figure \ref{fig:rsg_example} shows four example \Gaia light curves from our group (d) stars. 
The top row shows two typical supergiant light curves that are well sampled by the \Gaia time series.
In the bottom row, the left example shows a case that has very large variability amplitude and a light curve reminiscent of a Mira.
We therefore think that this star is a massive AGB star close to the end of its evolution, which is also supported by the presence of a bump in the light curve before maximum light \citep{mcsaveney2007}.
The very red colour of this object is remarkable considering that in \ks the star belongs to the most luminous group, while in \gmag it is of average brightness. 
The fourth example is a supergiant with a period that is too long to be fully covered by the time span included in DR2.
Such objects, although they come with a period in the DR2 catalogue, were excluded from our analysis.
We also rejected stars that had badly sampled or irregular light curves.
Excluding those objects we ended up with a total sample of 78 group (d) stars in the LMC and 17 in the SMC, which is in both cases about 45\% of all candidates in this group. 

\section{Results}
\label{sec:results}

Relations between period and luminosity are a common tool to study pulsational properties, such as the dominant pulsation mode, of a distinct group of variable stars.
In observational studies of LPVs, the luminosity, a quantity rarely available for large samples of stars, is typically replaced by the absolute $K$-band magnitude or the Wesenheit function \wjk.
In our study, we follow this approach using the $P$-\ks and $P$-\wjk diagrams, and further investigate the PLDs using two luminosity indicators derived from \Gaia, namely \wrp and \mbol.

The study was performed for both the LMC and SMC.
Assuming that all stars within one of the Clouds have the same distance, we did not apply distance corrections to the used magnitudes.
The average line-of-sight depths of the LMC and SMC were estimated by \citet{2009A&A...496..399S} to be about 4~kpc and ~4.5kpc, respectively, i.e. less than 10\% of their distance to the Sun.
These depths would translate into magnitude depths of 0.2~mag. 
However, for the SMC significantly larger depths for some directions occur \citep{ripepi_et_al_2017, sun_et_al_2018}. 
The line-of-sight depths are expected to add some scatter around the P-L relations, in particular for the SMC.
No reddening corrections were applied either, in accordance with the very low impact of interstellar extinction in the \ks\  band for objects outside the disc.
Besides, the small differential interstellar extinction across each Cloud \citep[e.g.][]{milone_etal_2009} leads to small dispersion of the \mbol values.

We start this section with a qualitative description in Sect.~\ref{sec:PLDs} of the PLDs using the four luminosity indicators \ks, \wjk, \wrp, and \mbol.
The description is then further detailed for intermediate- and high-mass O-rich stars in Sects.~\ref{Sect:IMstars} and \ref{Sect:HMstars}, respectively.
The case of C-stars is tackled in Sect.~\ref{Sect:Cstars}.

\subsection{Period-luminosity diagrams for various luminosity indicators}
\label{sec:PLDs}

\begin{figure*}
\centering
\includegraphics[width=\textwidth]{{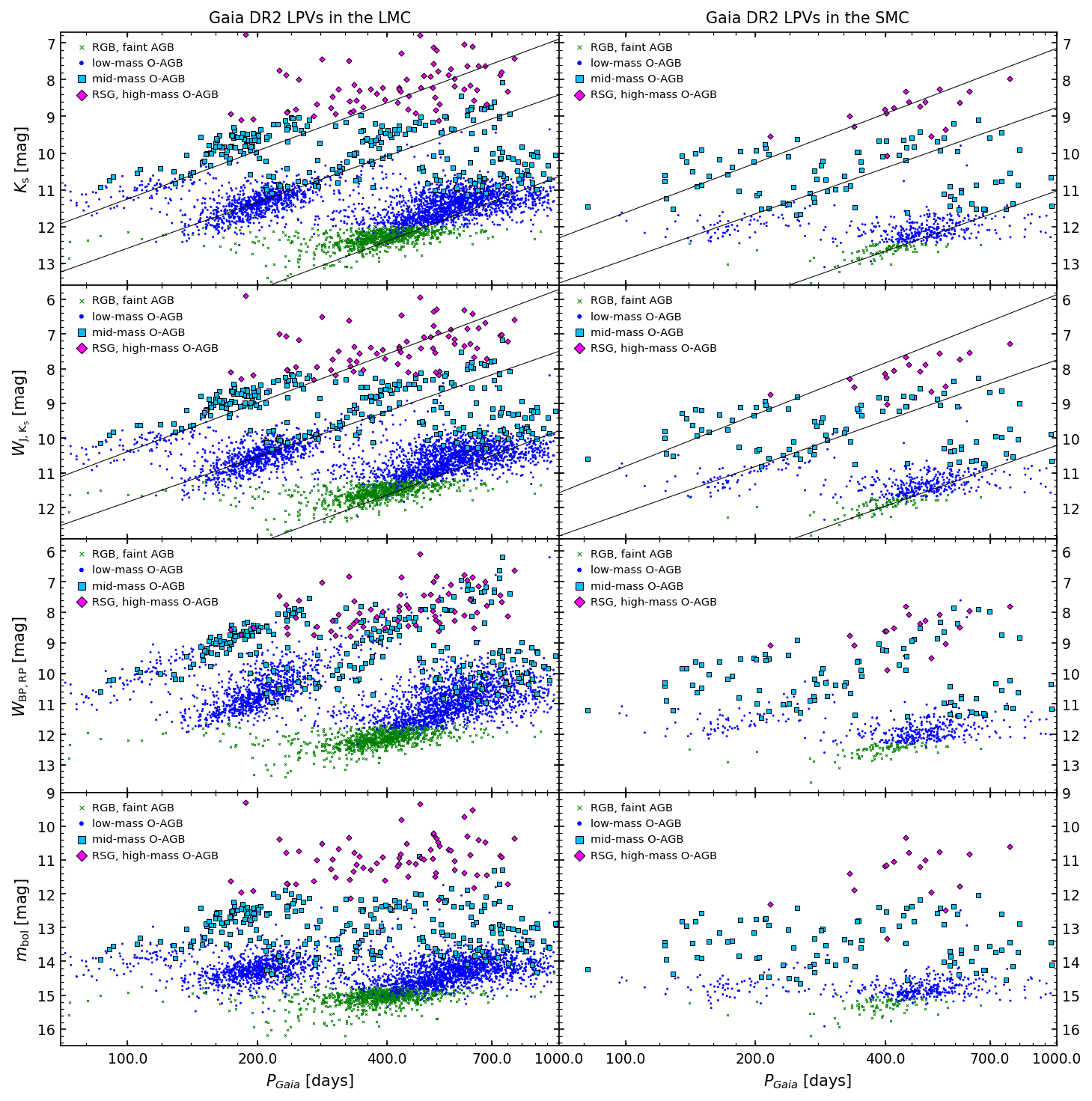}}
\caption{Period-luminosity diagrams of \Gaia LPV candidates in the LMC (left) and SMC (right), including only stars classified as O-rich, and using different luminosity indicators. From top to bottom,  $\ks$ band and $\wjk$ index from 2MASS, $\wrp$ index from \Gaia, and the bolometric magnitude computed as described in Appendix~\ref{sec:bc}. Lines in the top four panels are best fit to the P-L relations of O-rich LPVs from \citet{soszynski_etal_2007}.
        }
\label{fig:PLD_4x2_Orich}
\end{figure*}

\begin{figure*}
\centering
\includegraphics[width=\textwidth]{{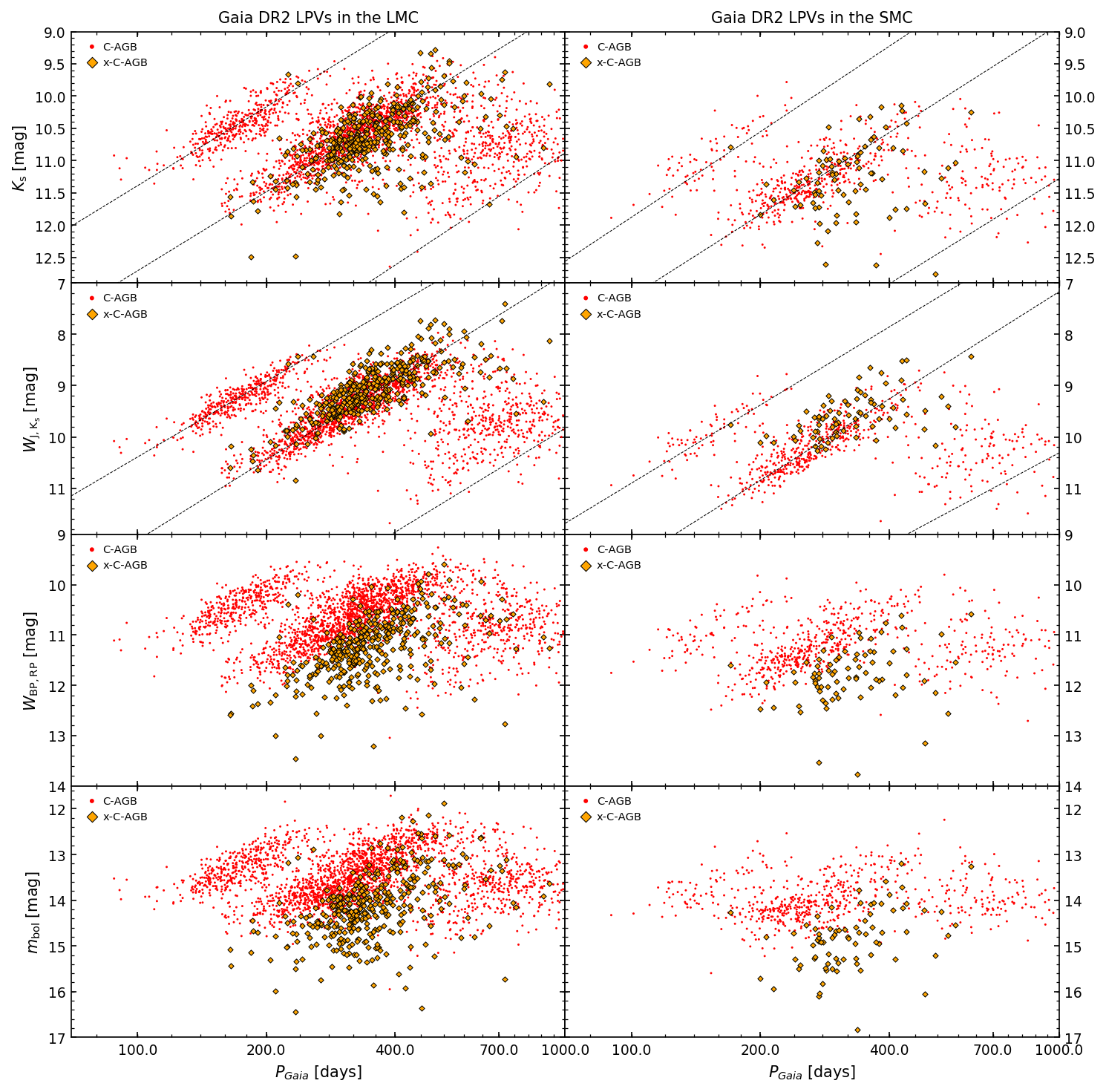}}
\caption{Period-luminosity diagrams of \Gaia LPV candidates in the LMC (left) and SMC (right), including only stars classified as C-rich (with extreme C-rich stars noted x-C-AGB in the panels), and using different luminosity indicators. From top to bottom, $\ks$ band and $\wjk$ index from
2MASS, $\wrp$ index from \Gaia, and the bolometric magnitude computed as described in Appendix~\ref{sec:bc}. Lines in the top four panels are best fit to the P-L relations of C-rich LPVs from \citet{soszynski_etal_2007}.
        }
\label{fig:PLD_4x2_Crich}
\end{figure*}

The P-\ks diagram of our complete samples for the two Clouds are shown in Fig.~\ref{fig:PLD_logP_Ks_LMC_SMC}.
The LPVs for which a period has been measured from \Gaia DR2 data populate the bright ends of sequences C$^{\prime}$, C, and D, as visible in the figure. 
We point out that the \Gaia data set extends the OGLE-III observations \citep{soszynski_etal_2009_ogle3lmc,soszynski_etal_2011_ogle3smc} towards higher brightness by almost 2\,mag in \ks.
Our study thus fills the gap between low-mass giants and supergiants left in previous studies \citep[see figure 7 of][]{kiss_etal_2006}.

The consistency with the best-fit relations derived by \citet{soszynski_etal_2007}, also shown in Fig.~\ref{fig:PLD_logP_Ks_LMC_SMC}, further testifies the good agreement with OGLE data discussed above. 
Sequence D as observed by \Gaia is more heavily scattered \citep[cf.][their figure~1]{soszynski_etal_2009_ogle3lmc} and systematically shifted towards shorter periods, a consequence of the degraded precision in the determination of periods longer than $\sim500$ days from DR2 data.
O-rich stars with masses $\lesssim 1.5\,{\rm M}_{\odot}$ [branch (a) in Fig.~\ref{fig:LMC_SMC_G2MD_apparent}] are observed predominantly on sequence D, meaning that their primary variability is associated with a LSP. 
This is especially true for stars in the (a-f) group in Fig.~\ref{fig:LMC_SMC_G2MD_apparent}, classified as RGBs or faint AGBs. 
These stars usually also pulsate in the 1O mode, but with a lower amplitude than the LSP and a relatively short period \citep[on the left side of sequence C$^{\prime}$, e.g.][]{trabucchi_etal_2017,wood_2015}, so that it is less likely characterised in DR2.

The PLDs defined by the other luminosity indicators, namely \wjk, \wrp, and \mbol, are shown in Fig.~\ref{fig:PLD_4x2_Orich} for O-rich stars, and in Fig.~\ref{fig:PLD_4x2_Crich} for C-rich stars.
In light of the all-sky coverage of the \Gaia survey, the use of \wrp and \mbol, which enables a study of PLDs solely based on data from this spacecraft, provides an important step in the full exploration of the \Gaia data set on LPVs.
Figures~\ref{fig:PLD_4x2_Orich} and \ref{fig:PLD_4x2_Crich} are organised in the same way, giving P-\ks, P-\wjk, P-\wrp, and P-\mbol from top to bottom, respectively.
The left column in each figure is used for LMC data, the right for SMC data.
The colour-coding of the various groups is explained in the top left corner of each panel.

All four luminosity indicators in Figs.\,\ref{fig:PLD_4x2_Orich} and \ref{fig:PLD_4x2_Crich} lead to a set of clear relations between the indicator and the period.
At the same time we immediately note significant differences between the four kinds of diagrams and between the relations for the various groups of LPVs identified in Fig.~\ref{fig:LMC_SMC_G2MD_apparent}.

For the O-rich stars (C-rich stars are discussed in Sect.~\ref{Sect:Cstars}), we note the following points when comparing the diagrams using the different luminosity indicators (see Fig.~\ref{fig:PLD_4x2_Orich}):
(1)The P-\ks- and the P-\wjk diagrams show a high similarity as has been outlined in the literature before \citep[e.g.][]{2013ApJ...763..103S}.
(2) In the P-\wrp diagram (third row in Fig.~\ref{fig:PLD_4x2_Orich}), sequence C' is narrower and thus seemingly better defined compared to the infrared indicators.
We note that opposite to \wjk the quantity \wrp consists of photometry averaged over a light cycle. 
This may in part be relevant for the smaller scatter.
However, only 5\% of the stars on C' have large amplitudes in \gmag ($>$1 mag), and they are expected to have even smaller amplitudes in \ks. 
Therefore, the effect of averaging is expected to be comparably small.
This also suggests that differences between the widths of the sequences in various luminosity indicators are not primarily caused by using single or averaged multi-epoch photometry.
(3) The structures seen in the P-\mbol diagram (bottom row of Fig.~\ref{fig:PLD_4x2_Orich}) coincide more with the infrared luminosity indicators than with \wrp. 
While the diagram still facilitates a distinction between various groups, relations, in particular for the low-mass stars, are flatter.
A difference in inclination between the relations for low- and intermediate-mass stars is likely present.
The supergiants and the high-mass AGB stars do not show any distinct sequences in this diagram.
The findings described in this section are applicable to both the LMC and SMC sample, although the SMC sample is less clear because of the lower number of stars.

\subsection{Intermediate-mass AGB stars.}
\label{Sect:IMstars}

A large fraction of the bright AGB stars in the \Gaia sample, which \citet{lebzelter_etal_2018} classified as intermediate-mass O-rich AGBs [branch (c)] or as RSGs and massive AGBs [branch (d)], are not present in the OGLE-III database, and are thus of particular interest. 
Stars on branch (c) clearly follow three P-L relations, which is consistent with sequences C$^{\prime}$, C, and D. But these are systematically shifted towards shorter periods compared to low-mass O-rich stars in the infrared PLDs (first and
second rows in Fig.~\ref{fig:PLD_4x2_Orich}).
This is a consequence of their higher masses and agrees with the theoretical prediction of \cite{wood_2015}. 
This offset is not visible when the $\wrp$ index is employed (Fig.~\ref{fig:PLD_4x2_Orich}, third row), and its absence accounts for a much narrower sequence C' compared to \ks and \wjk. 
In the P-\mbol diagram, the offset between the relations for low- and intermediate-mass LPVs is clearly detectable only for sequence C'. 
Sequence C is very wide for the intermediate-mass stars and the location of the relation is much less defined than for \wjk, for example. 
However, a smooth transition between low- and intermediate-mass objects is suggested.
We explore the cause of this difference between the various luminosity indicators in Sect. \ref{Discus:PLD_Mass}.

\subsection{High-mass AGB and RSG stars.}
\label{Sect:HMstars}

The LPVs that belong to branch (d) have even higher masses than those on branch (c). 
Therefore, we would expect to find an offset for these objects as well.
However, their distribution in the PLD suffers from a larger scatter, and it is less straightforward to identify the P-L relation(s) they belong to. 
This is further complicated by the fundamentally different behaviour they exhibit depending on whether visual or near-infrared bands are used to track their luminosity. 
Judging by the PLDs in the upper panels of  Fig.~\ref{fig:PLD_4x2_Orich}, where $\ks$ and $\wjk$ are used, the bulk of stars from branch (d) appear to lie on a prolongation of sequence C$^{\prime}$. 
In contrast, when \Gaia photometry is used in the PLD (i.e. the $\wrp$ index, Fig.~\ref{fig:PLD_4x2_Orich}, third row), most of these bright stars seem to follow the P-L sequence C.

The P-\mbol diagram (fourth row in Fig.~\ref{fig:PLD_4x2_Orich}) does not show pronounced indications for a connection of  the high-mass LPVs and red supergiants to any of the relations found from low- and intermediate-mass stars.
The location of these stars in this diagram, however, is found in excellent agreement with the same kind of diagram presented by \citet{wood_etal_1983}; see their figures~6 and~7. 
The diagram of \citet{wood_etal_1983} included supergiants not only from the Magellanic Clouds but also from the Galactic field. 

There are no high-mass O-rich LPVs that show a variation connected to sequence D.
The existence of such LSPs had been reported by several authors \citep[e.g.][]{kiss_etal_2006}.
However, we determined only a single period and the corresponding sequence D period would likely exceed our upper period cut-off.

\subsection{C-stars}
\label{Sect:Cstars}

The P-L relations of C-rich stars [groups (b) and (b-x)] are shown in Fig.~\ref{fig:PLD_4x2_Crich}. 
For these objects, the differences in the PLD arising from using distinct luminosity indicators become even more striking. 
When using \wrp (third row of Fig.~\ref{fig:PLD_4x2_Crich}), the highly reddened C-stars are shifted away from the bluer C-stars and the M-stars (Fig.~\ref{fig:PLD_logP_Ks_LMC_SMC}) towards lower brightness.
This indicates that the \Gaia Wesenheit function \wrp does not compensate sufficiently the reddening of large amounts of circumstellar dust affecting \grp.
We do not observe such a difference in \wjk (upper panels of Fig.~\ref{fig:PLD_4x2_Crich}), indicating that $\gbp-\grp$ gets saturated for large extinction values opposite to $J-K$, where the thick circumstellar shell starts to emit in the infrared adding to the brightness in \ks and \wjk.
This agrees with results from the preliminary analysis of the \Gaia Wesenheit function presented in \citet{lebzelter_etal_2018}.
Since the BC used to compute \mbol also depends on $\gbp-\grp$, we observe the same effect as for \wrp in the $P$-\mbol diagram (bottom panels in Fig.~\ref{fig:PLD_4x2_Crich}).

Aside from the offset of the highly reddened C-stars, the P-\wrp and P-\mbol diagrams show structures very similar to the P-\ks and P-\wjk diagrams. 
Like the O-rich LPVs, the C-rich LPVs [branch (b)] follow all three P-L relations, while the extreme C-rich AGBs [group (b-x)] are likely to follow only sequence C, as is most evident when using $\wjk$ in the PLD (Fig.~\ref{fig:PLD_4x2_Crich}, second row). 
The few stars of this group that are found on sequences C$^{\prime}$ or D are very close to the separation line between C-rich and extreme C-rich stars in Fig.~\ref{fig:LMC_SMC_G2MD_apparent}. 
Therefore, a misclassification cannot be excluded. 
Those extreme C-stars below sequence C probably have very thick shells (i.e. high mass-loss rates) and emit mostly in the far-infrared so \ks is indeed faint. 
Again, a qualitatively similar behaviour is seen in both the LMC and SMC.


It is noteworthy that C-stars forming the P-L sequence with the longest periods, usually dubbed sequence D, is clearly offset to the location of sequence D defined by carbon stars in \citet{soszynski_etal_2007}, shown by the dotted lines in Fig.~\ref{fig:PLD_4x2_Crich}.
We found a smaller shift in the same direction when studying the O-rich stars (Fig.~\ref{fig:PLD_4x2_Orich}), which is likely caused by the limited time span of the \Gaia light curves and the upper period limit resulting from it. 
This effect is more expressed for C-stars than for O-rich objects, likely because of their on average longer periods compared to M-type stars.
Having a period closer to our period cut-off, there is a higher risk that our period search fails to derive the correct value.

\section{Discussion} \label{Sect:Discussion}

\subsection{Period-luminosity relations at different masses}
\label{Discus:PLD_Mass}

As illustrated in the top two rows of Fig.~\ref{fig:PLD_4x2_Orich} and as pointed out in Sect.\,\ref{Sect:IMstars}, there is a very obvious shift between the P-L relations of low- and intermediate-mass O-rich stars; the latter show a shorter period at a given luminosity. This shift has 
previously been noted for sequence C (the Miras) by \citet{feast_et_al_1989} and \citet{hughes_wood_1990}.
As briefly mentioned above, such a shift is expected to occur because a higher mass of the pulsating red giant leads to a shorter period at a given luminosity. 
\citet{whitelock_etal_2003} noted that at least some of the stars shifted to a shorter period from sequence C had spectral signatures of hot-bottom burning, consistent with their intermediate-mass assignment.  
This is the first time the shift in period with mass has been observed for LPVs pulsating in the 1O.
Because of the conversion of stars in the mass range between groups (a) and (c) into carbon stars, we do not see a smooth transition but rather a break up into two sequences.

\begin{figure*}
\centering
\includegraphics[width=\textwidth]{{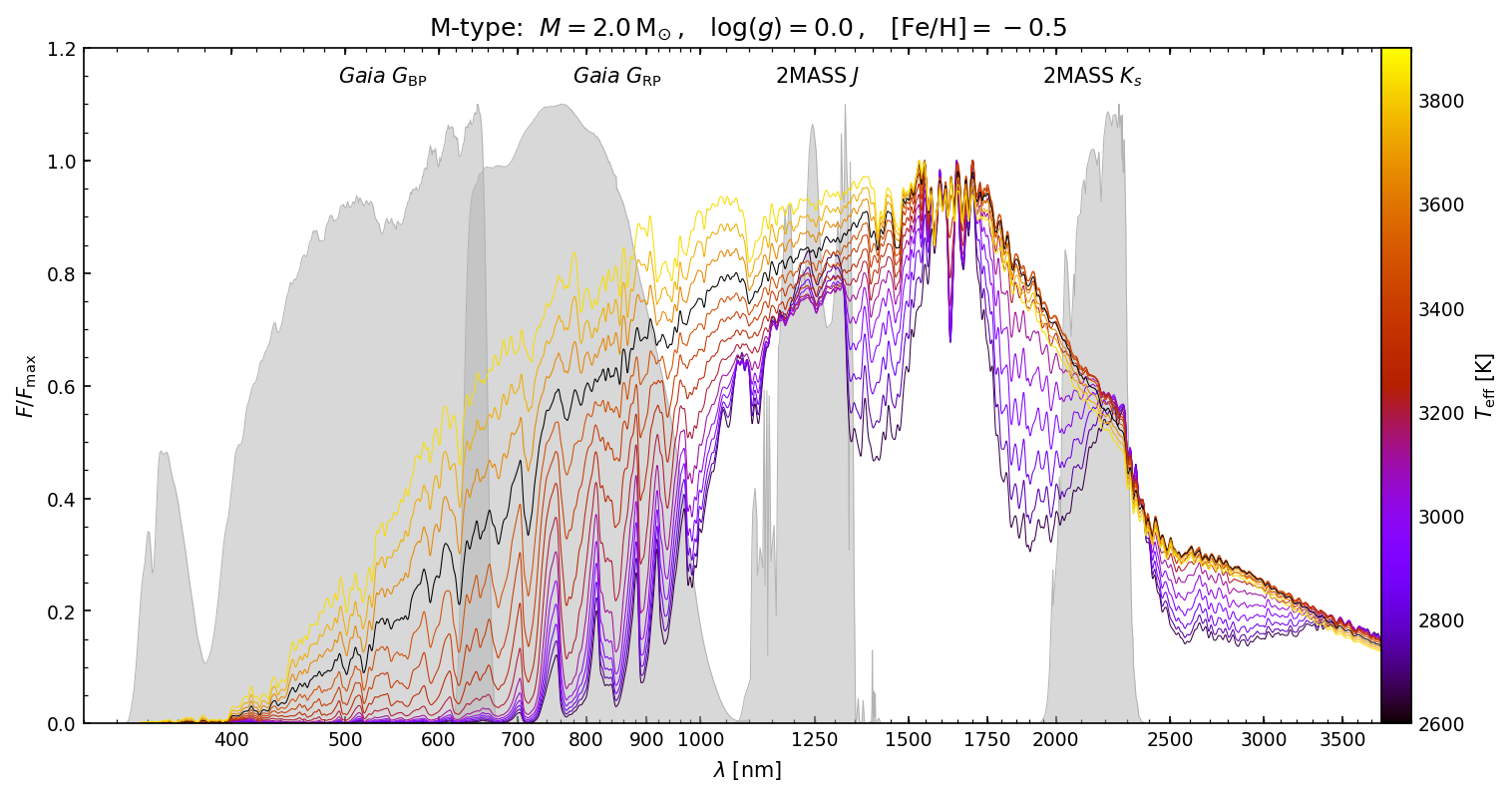}}
\caption{Synthetic spectra \citep{aringer_etal_2016} of O-rich stars with $M=2\,\Msun$, $\log(g)=0.0$ and different effective temperatures. The black line indicates the model with $T_{\rm eff}=3600\,{\rm K}$.
        }
\label{fig:spectrum_and_bands}
\end{figure*}

\begin{figure*}
\centering
\includegraphics[width=\textwidth]{{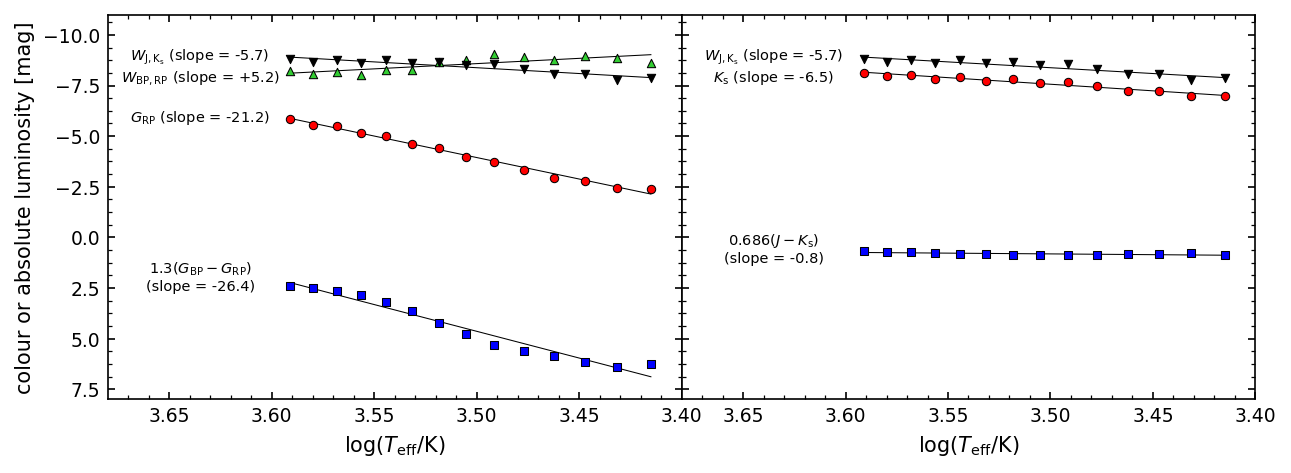}}
\caption{Sensitivity on temperature of the various filters and filter combinations used in the present analysis. Values were derived using synthetic spectra of O-rich stars with $M=2\,\Msun$, $\log(g)=0.0$ and different effective temperatures (cf.\,Fig.~\ref{fig:spectrum_and_bands}).
        }
\label{fig:Tdepend_filters}
\end{figure*}

Interestingly, the shift disappears in the P-\wrp diagram. 
To understand this, we have to keep in mind that a star of higher mass also shows a higher effective temperature, which again may affect fluxes in the various photometric bands and accordingly also colours.
We explored the effect of temperature on the various filters with the help of hydrostatic models from \citet{aringer_etal_2016}. 
Fig.~\ref{fig:spectrum_and_bands} presents model spectra of O-rich AGB stars of various temperatures, and compares them with the responses of the two \Gaia and the two near-infrared passbands used in our analysis.
 Figure~\ref{fig:Tdepend_filters} shows the runs of \ks, $J-K_s$, \wjk, \gbp-\grp, and \wrp with temperature; 
\ks and $J-K_s$ are only weakly dependent on temperature.
As a consequence, \wjk behaves very similarly to \ks, and any systematic temperature difference between low- and intermediate-mass stars does not become visible in the corresponding PLDs. 

In the \gbp filter, the absorption bands of TiO make a huge impact on the flux, which therefore is highly dependent on the temperature of the star. 
Since the \grp flux is less affected by TiO and VO absorption than the \gbp flux, the colour constructed from these two filters shows a high sensitivity on temperature, in particular down to about 3000\,K.
To compute \wrp, the temperature-sensitive colour \gbp-\grp is subtracted from the \grp, thereby compensating for the flux depression in \grp.
However, we see in Fig.~\ref{fig:Tdepend_filters} that while \grp and \gbp-\grp are decreasing in value with temperature, the combined quantity \wrp is increasing in value towards higher temperatures because the slope of   the temperature dependency of 1.3(\gbp-\grp) is steeper than for \grp.
The factor 1.3 used in the construction of the Wesenheit function obviously leads to an overcompensation of the temperature dependencies of the individual \Gaia filters. 
Both intermediate-mass and low-mass LPVs are made brighter in \wrp, but this occurs to a smaller extent for the former owing to their higher temperature.
The two effects, the offset due to mass (i.e. shorter period due to higher mass at given luminosity) and the effect of temperature on \wrp, equal out, and thus sequences C' and C become single sequences more narrow than their near-infrared counterparts.

When interpreting the P-\mbol diagram in this context, we have to be careful since the BC is a function of \gbp-\grp. 
Therefore, our \mbol values are expected to exhibit some similarity with the \wrp values, although the factors in the colour terms involved are very different. 
As a consequence, the P-\mbol sequences of low- and intermediate-mass stars do not align as for P-\wrp.

\begin{figure}
\centering
\includegraphics[width=\columnwidth]{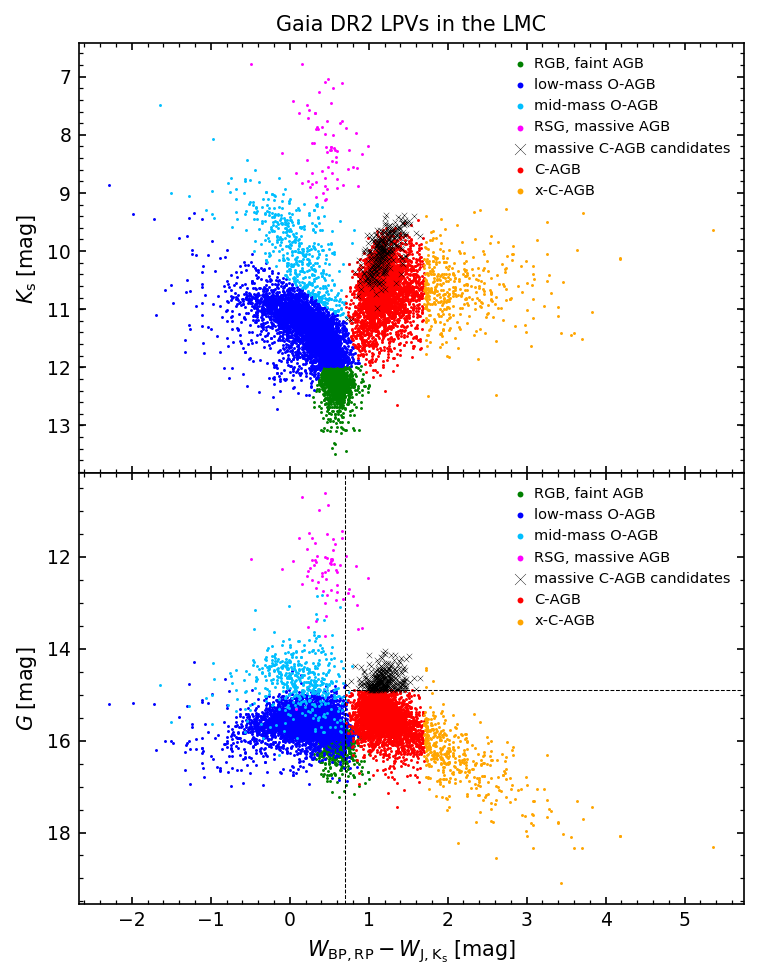}
\caption{Top panel: Same as Fig.~\ref{fig:LMC_SMC_G2MD_apparent}, but massive C-star candidates are highlighted as black crosses. Bottom panel: Selection of massive C-star candidates in a modified \Gaia-2MASS diagram using the \Gaia $G$ band on the vertical axis.}
\label{fig:massiveCcandidatesG2MD}
\end{figure}

\begin{figure}
\centering
\includegraphics[width=\columnwidth]{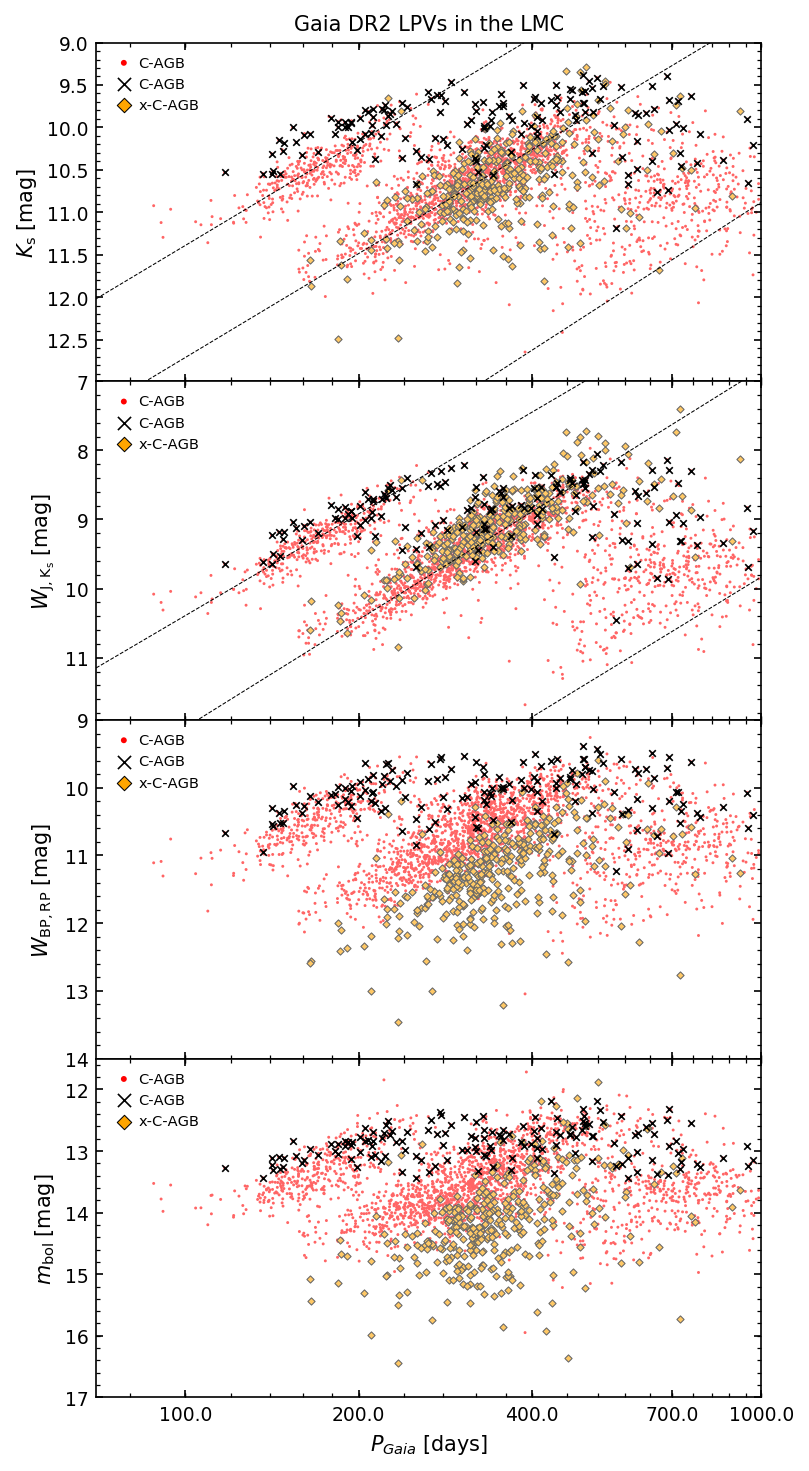}
\caption{Same as Fig.~\ref{fig:PLD_4x2_Crich}, with massive C-star candidates highlighted as black crosses.}
\label{fig:massiveCcandidatesPeriods}
\end{figure}

After detecting the mass dependency of the P-L relation for O-rich stars in our observations, we decided to investigate the C-rich stars for indications of a mass dependency as well.
With the onset of efficient dust production, circumstellar absorption systematically depresses brightness in all optical and near-infrared filter bands. 
Therefore, such an investigation has to be limited to C-stars with low amounts of dust around them, i.e. group (b) stars in the \Gaia-2MASS diagram.
There is no doubt that with this approach some of the most massive C-stars are excluded.
Within group (b) in our diagram, a separation of the stars according to mass is not obvious.
Using the brightest C-stars in \ks gives a sample that is scattered over the bright part of the PLDs in Fig.~\ref{fig:PLD_4x2_Crich}. 
However, selecting the brightest C-stars in the \gmag\  band (bottom panel of Fig.~\ref{fig:massiveCcandidatesG2MD}) gives a sample of stars that form a sequence in the P-\ks diagram slightly offset to the bulk of carbon stars on sequence C'
(Fig.~\ref{fig:massiveCcandidatesPeriods}). 
This group strongly recalls the behaviour of the intermediate-mass stars in the O-rich case.
For C-stars, the luminosity indicator \wjk gives the most narrow sequences.
We suspect, in analogy to the O-rich stars, that the C-stars delineated in Fig.~\ref{fig:massiveCcandidatesPeriods} are more massive than the majority of the group (b) C-stars in our sample, and that \wjk, like \wrp in the O-rich case, compensates for the temperature and mass dispersion leading to single, narrow sequences C' and C. 

\subsection{Comparison with evolutionary models for low- and intermediate-mass stars}

\begin{figure*}
\centering
\includegraphics[width=\textwidth]{{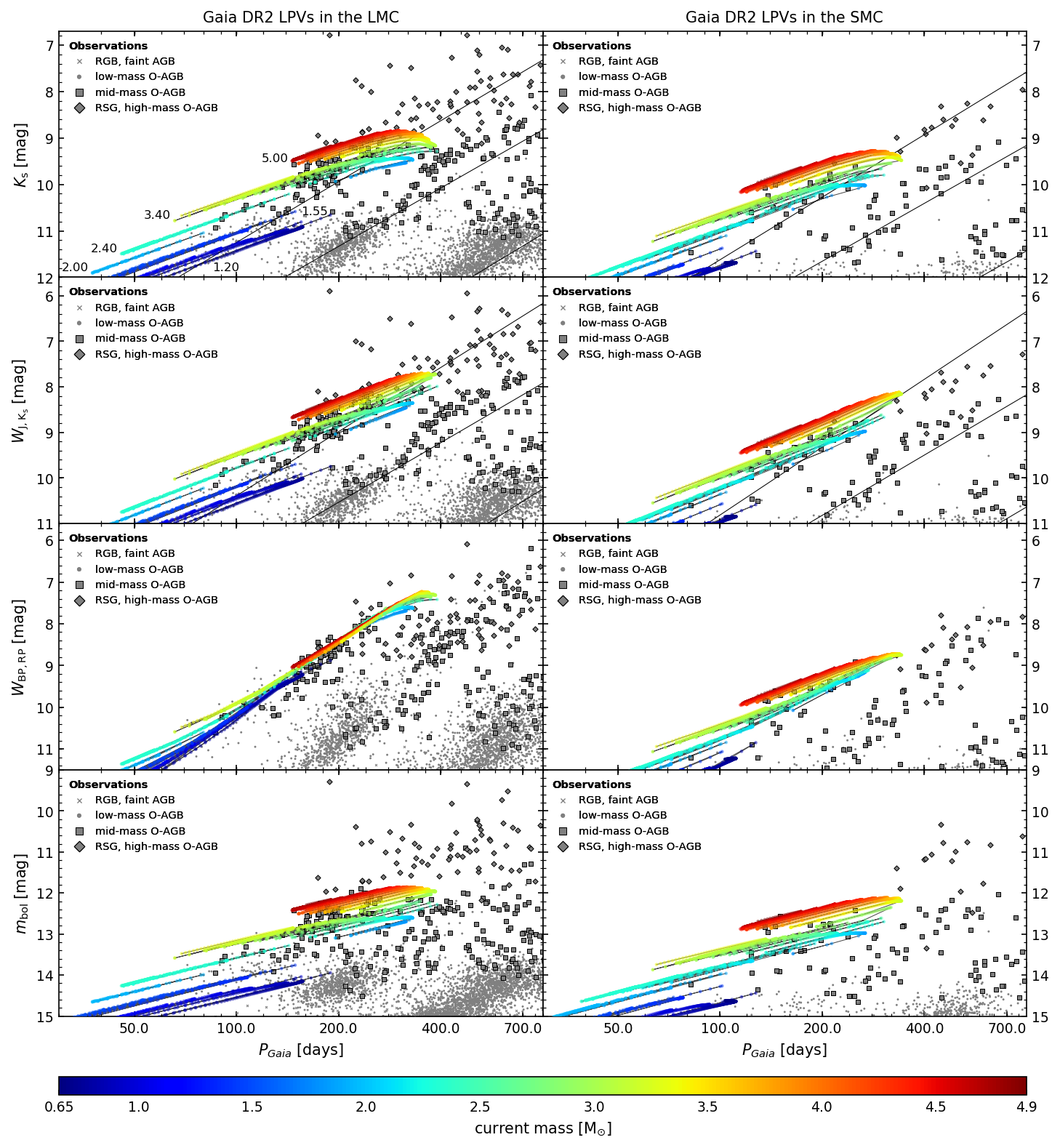}}
\caption{Same as Fig.~\ref{fig:PLD_4x2_Orich}, but including AGB models with 1O mode periods. Evolutionary tracks are limited to the portion where ${\rm C/O}<1$ and the 1O mode is expected to be dominant. Tracks are colour coded with current mass, and labelled with their initial mass in the top left panel. They have initial metallicity Z=0.006 (left panels) and Z=0.002 (right panels).
        }
\label{fig:PLD_4x2_Orich_wModels}
\end{figure*}

\begin{figure*}
\centering
\includegraphics[width=\textwidth]{{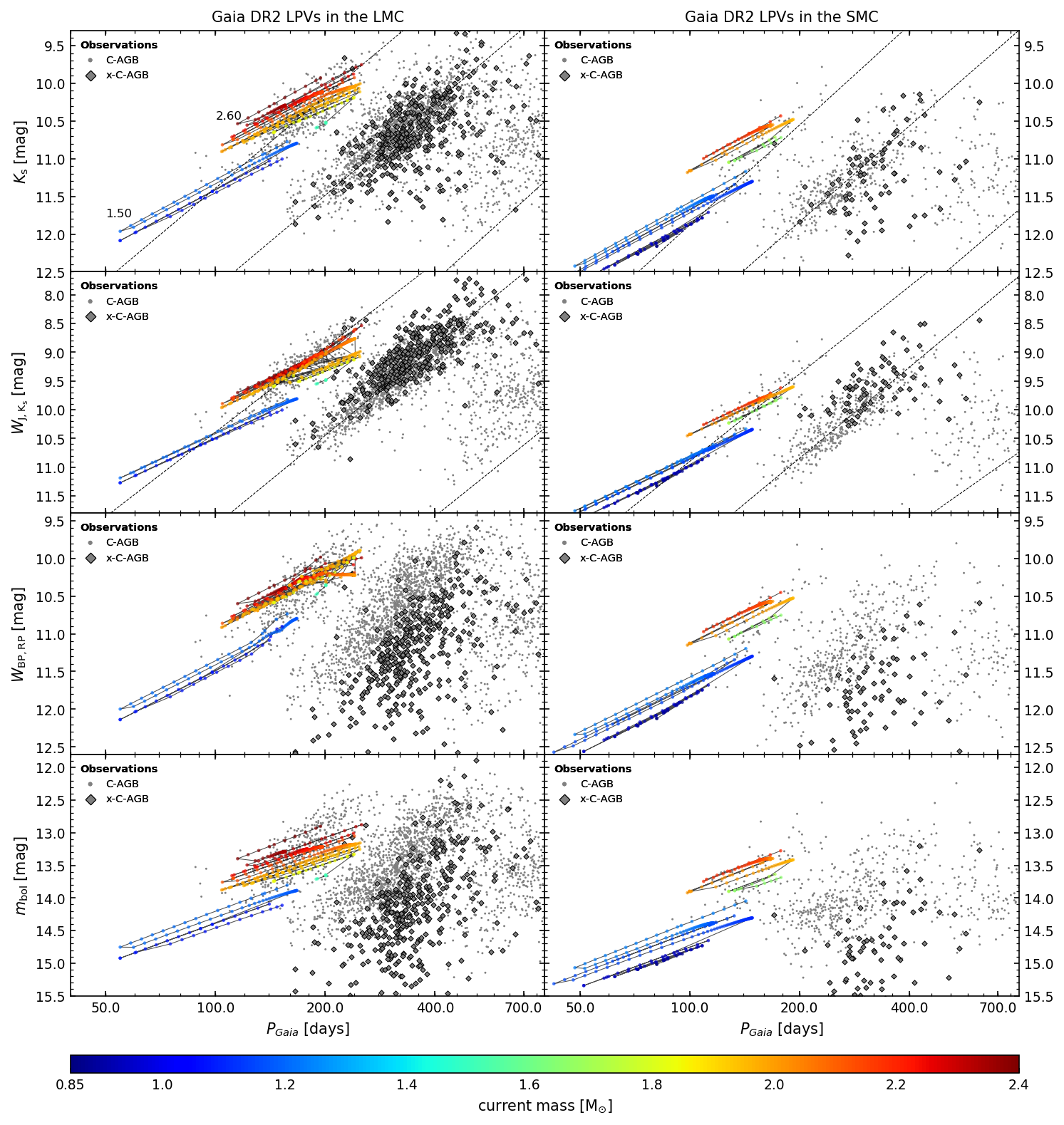}}
\caption{Same as Fig.~\ref{fig:PLD_4x2_Orich_wModels}, but comparing observed stars on branch (b) with the portions of evolutionary tracks having ${\rm C/O}>1$. The represented tracks roughly correspond to the minimum and maximum initial masses that produce carbon stars pulsating in the 1O mode at the chosen metallicities. For clarity, tracks with an initial mass that is intermediate between these two values have been omitted.
        }
\label{fig:PLD_4x2_Crich_wModels}
\end{figure*}

We compare the observed data of LPVs with the AGB evolutionary tracks computed with the \texttt{COLIBRI} code \citep{marigo_etal_2013} and presented in \citet{pastorelli_etal_2019}. 
The models used for the present work cover the mass range between 1 and 5\,\Msun for two choices of the initial metallicity, Z=0.006 and Z=0.002. 
Along each AGB track periods and amplitude growth rates are calculated using the pulsation models of \citet{trabucchi_etal_2019}. 
The results for 1O mode models are shown in Figs.~\ref{fig:PLD_4x2_Orich_wModels} and \ref{fig:PLD_4x2_Crich_wModels}.
For the SMC, the selected tracks refer to  $Z = 0.002$, thus slightly underestimating the average metallicity of this galaxy.
This has to be kept in mind in the comparison of models and observations.

The colour coding represents the current mass of the star, hence there are colour changes along the tracks.
Current pulsation models of LPVs suffer from a known tendency to overestimate the period of the fundamental mode, especially at large luminosities \citep[e.g.][]{trabucchi_etal_2017}, hence we preferred to limit the comparison to 1O mode periods, for which we find a good agreement with observations.

Features identified in the four kinds of PLDs are well reproduced by our set of models.
The `mass spread' across individual P-L sequences is clearly visible in the plots of period versus \ks and \wjk. 
When periods are shown against \wrp, this effect is suppressed, which is clearly visible for LMC metallicities ($Z\simeq0.006$) and is less obvious in the case of the SMC ($Z\simeq0.002$).
To understand the difference between the two Magellanic Clouds we need to consider that the lower metallicity leads to weaker molecular bands at a given temperature, which again reduces the sensitivity of \gbp-\grp on temperature.
In contrast to the more metal-rich case of the LMC, where the colour term in the Wesenheit function compensates for temperature and mass shift, we see --  in the observations and models -- a slight offset between the P-\wrp-relations of low- and intermediate-mass stars, respectively.
Overall, the model results agree well with our observational findings, supporting the discussion presented in Sect.\,\ref{Discus:PLD_Mass}.

For the C-rich LPVs, Fig.~\ref{fig:PLD_4x2_Crich_wModels}, the calculated tracks for 1O pulsation show the same behaviour as the observations when going from a P-\ks diagram (first row) to a P-\wjk diagram (second row), namely a narrowing of sequence C'. 
We emphasise that in this diagram the current masses of the model stars are shown, corresponding to main sequence masses between 1.5 and 2.6\,\Msun.
We decided to show the current mass since it is relevant for the pulsation properties of the star.
The flattening of the sequences seen in the P--\mbol diagram is reproduced by the models as well.

\begin{figure*}
\centering
\includegraphics[width=\textwidth]{{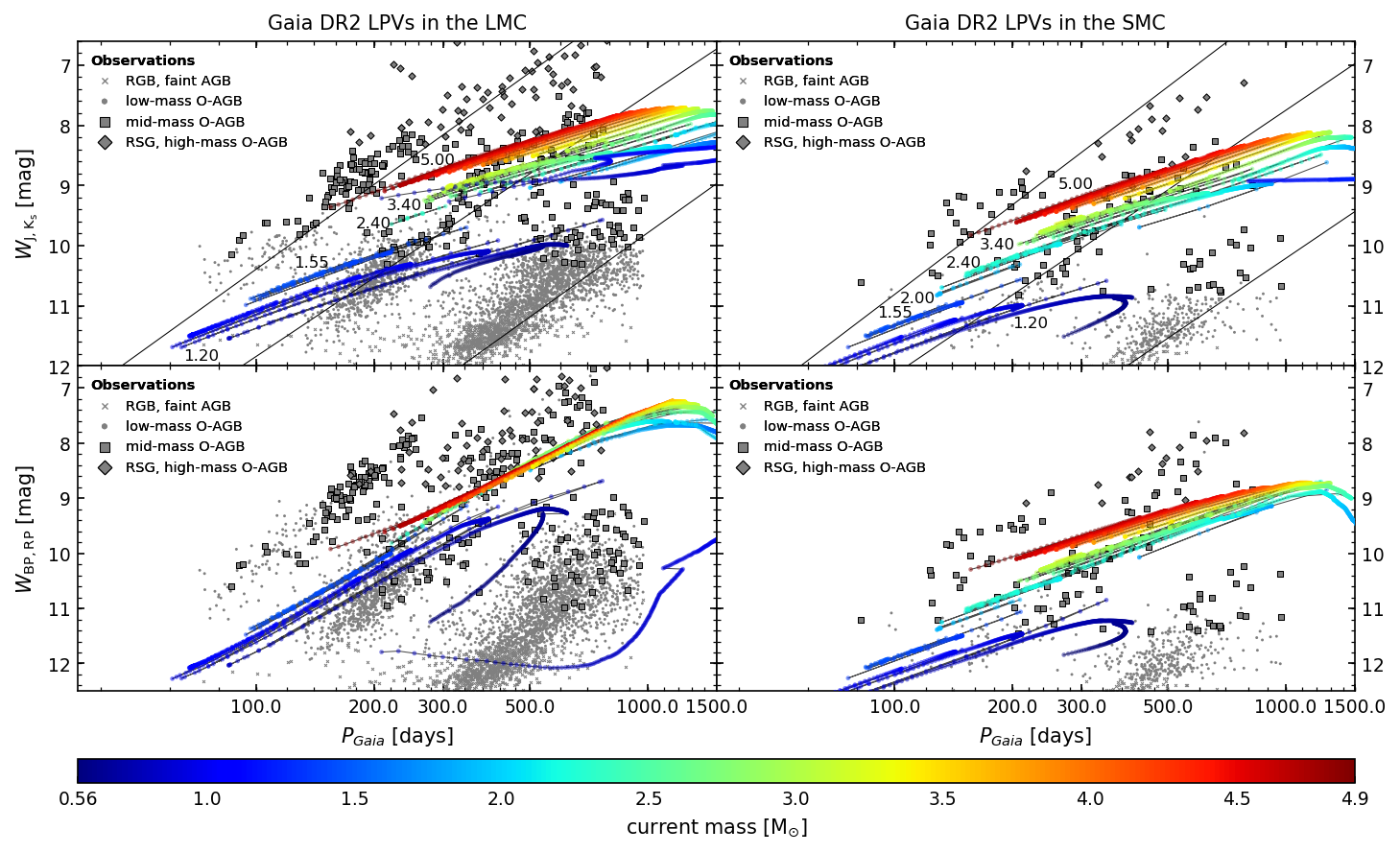}}
\caption{Same as Fig.~\ref{fig:PLD_4x2_Orich_wModels}, but showing predicted fundamental mode periods along evolutionary tracks.
        }
\label{fig:PLD_4x2_Orich_wModelsF}
\end{figure*}

For completeness, results for fundamental mode pulsation of O-rich models are shown in Fig.~\ref{fig:PLD_4x2_Orich_wModelsF} for the two luminosity indicators \wjk and \wrp.
Keeping in mind the limitations of the pulsation models for predicting fundamental mode periods, an overall agreement with the observations is present.
Periods become visibly too long at bright magnitudes.
The observed narrowing of sequence C is well reproduced by stellar evolution and pulsation models. 

\begin{figure*}
\centering
\includegraphics[width=\textwidth]{{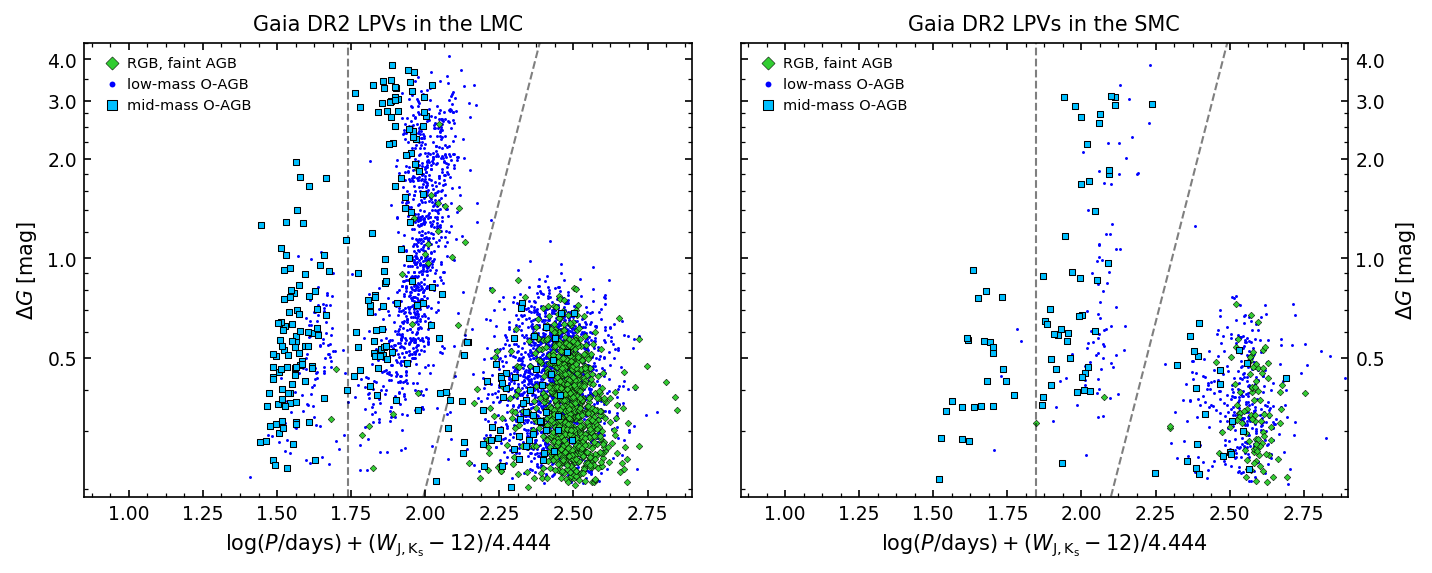}}
\caption{Projected-period vs. \Gaia G-band amplitude (see text) of low- and intermediate-mass O-rich AGBs in the LMC (left) and SMC (right). The green diamonds indicate RGBs and faint AGBs; the blue symbols indicate low-mass O-rich AGBs; and the cyan squares indicate intermediate-mass O-rich AGBs. The dashed lines represent approximate separations between the three distributions corresponding to 1O mode periods, fundamental mode periods, and LSPs.
        }
\label{fig:PALD_LMC_SMC_Orich}
\end{figure*}

\begin{figure*}
\centering
\includegraphics[width=\textwidth]{{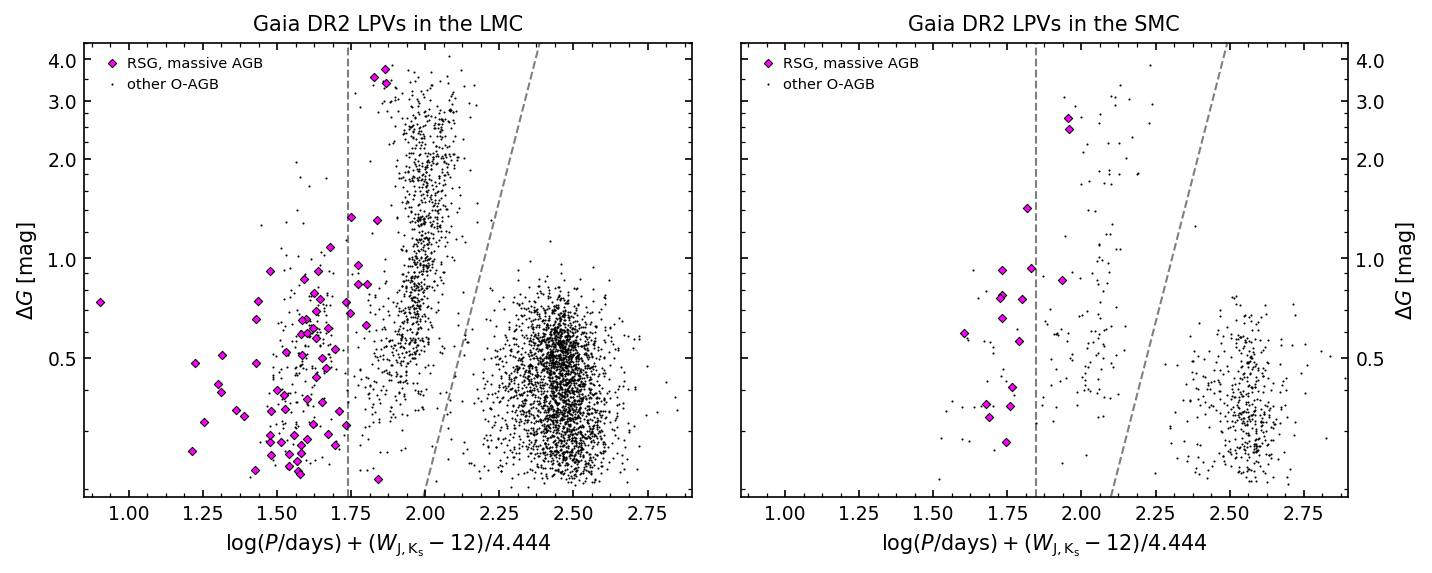}}
\caption{Same as Fig.~\ref{fig:PALD_LMC_SMC_Orich}, including the group consisting of RSGs and massive O-rich AGBs. Other O-rich AGB stars are shown as small black symbols, for reference. The dashed lines have the same meaning as in Fig.~\ref{fig:PALD_LMC_SMC_Orich}.
        }
\label{fig:PALD_LMC_SMC_RSG}
\end{figure*}

\begin{figure*}
\centering
\includegraphics[width=\textwidth]{{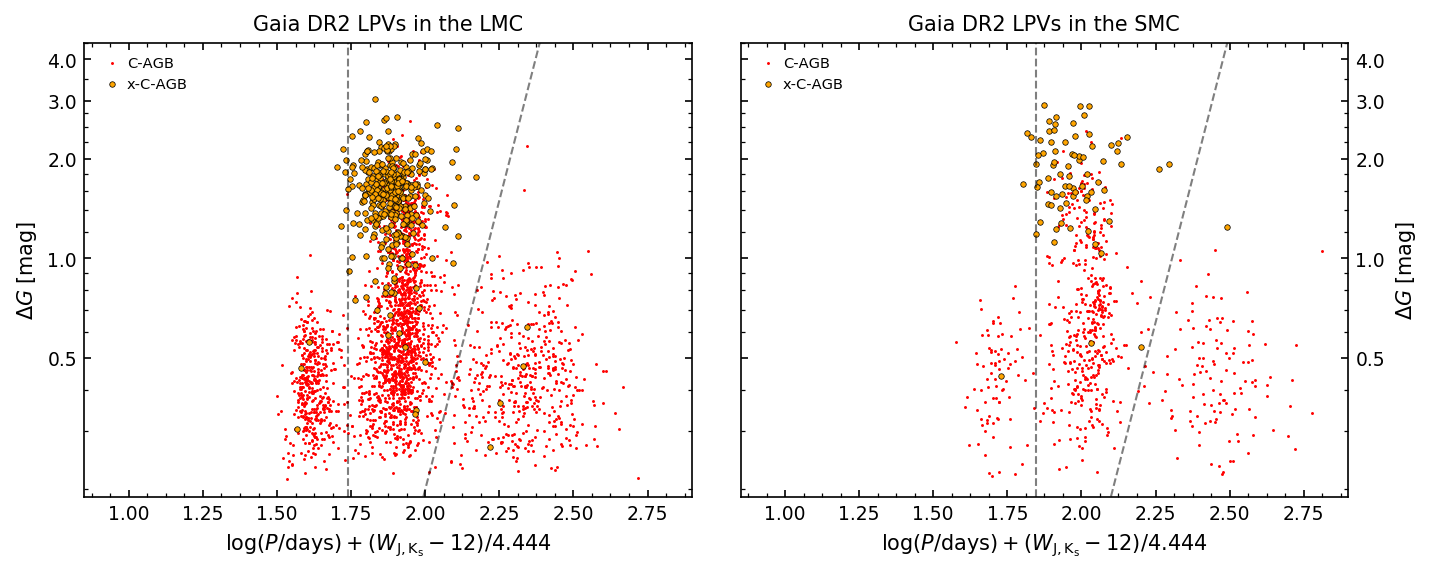}}
\caption{Same as Fig.~\ref{fig:PALD_LMC_SMC_Orich}, but for C-rich (red) and extreme C-rich (large orange symbols) AGB stars. The dashed lines have the same meaning as in Fig.~\ref{fig:PALD_LMC_SMC_Orich}.
        }
\label{fig:PALD_LMC_SMC_Crich}
\end{figure*}

Figures~\ref{fig:PALD_LMC_SMC_Orich}, \ref{fig:PALD_LMC_SMC_RSG}, and~\ref{fig:PALD_LMC_SMC_Crich} show a variant of the period-amplitude diagram, where the period on the horizontal axis is replaced by the quantity $\log(P)-(W_{\rm JK}-12)/\Delta_{P,W_{JK}}$ \citep{wood_2015}, where $\Delta_{P,W_{JK}}$=-4.444 is assumed to be a representative value for the average slope of the sequences in the $\log(P)$--$W_{JK}$ PLD. 
As shown by \citet{trabucchi_etal_2017}, this is a better tool than the usual PLD to discriminate between pulsation in different modes, particularly between fundamental mode pulsation (sequence C) and LSPs (sequence D). 
We point out that this study is limited to the period range 70 to 1000~d.

The three distributions visible in Fig.~\ref{fig:PALD_LMC_SMC_Orich} are periods associated with the 1O mode, the fundamental mode, and LSP variability \citep{trabucchi_etal_2017}. 
In both the LMC and SMC, the majority of stars classified as RGBs or faint AGBs show variability in a LSP. 
Less than 2 percent of LMC stars in this group show variability in the fundamental mode, and only two show variability in the 1O mode, while in the SMC only two stars of this group show 
fundamental or 1O variability compared to 91 stars that exhibit a LSP.
We note that such small numbers are highly sensitive to how the boundaries are defined in the \Gaia-2MASS diagram and to the chosen values of the distance moduli and the $\ks$-band luminosities of the RGB tip in the two systems. Low- and intermediate-mass O-rich AGBs have periods in all three groups. 
In contrast, the periods of RSGs and massive AGB stars systematically populate the region associated with 1O mode pulsation (except for a few large amplitude objects that pulsate in the fundamental mode) and exhibit no LSPs (Fig.~\ref{fig:PALD_LMC_SMC_RSG}).  We note, however, that the identification of pulsation modes in this diagram by \citet{trabucchi_etal_2017} is based on AGB pulsation models, and may not be valid for RSG variables (see Sect.\,\ref{Discus:modeRSG}). 
Finally, C-rich stars (Fig.~\ref{fig:PALD_LMC_SMC_Crich}) appear in all three groups, but extreme C-rich stars are essentially limited to pulsation in the fundamental mode, which is consistent with the fact that these are evolved objects with high mass-loss rates \citep{2009AJ....137.3139V}.

\begin{figure*}
\centering
\includegraphics[width=\textwidth]{{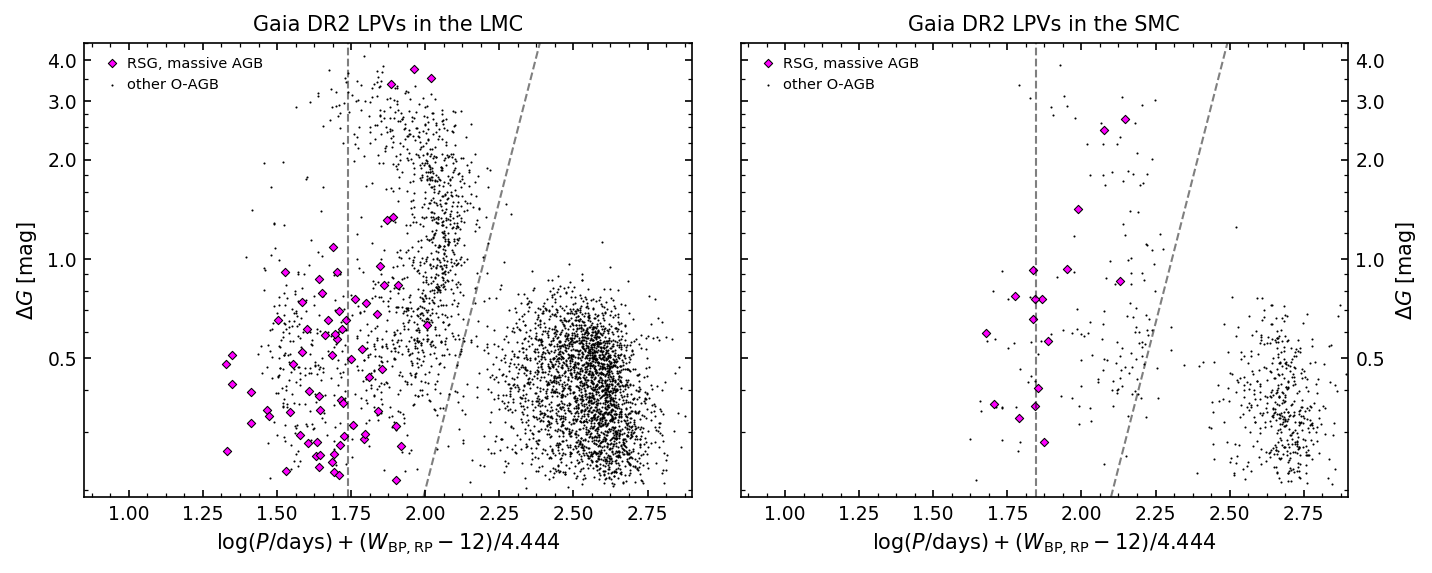}}
\caption{Same as Fig.~\ref{fig:PALD_LMC_SMC_RSG}, but the horizontal coordinate is obtained by projecting the period in the $\log(P)$ vs. $\wrp$ diagram, allowing us to better distinguish 1O mode periods from fundamental mode ones. The dashed lines have the same meaning as in Fig.~\ref{fig:PALD_LMC_SMC_Orich}.
        }
\label{fig:PALD_WRP_LMC_SMC_RSG}
\end{figure*}

\subsection{Pulsation mode of high-mass LPVs}
\label{Discus:modeRSG}

As noted in Sect.\,\ref{Sect:HMstars} these objects appear connected to different relations depending on the luminosity indicator used
(Fig.~\ref{fig:PLD_4x2_Orich}).
Taking the result from the P-\wrp diagram at face value would indicate that the bulk of the supergiants and massive AGBs are pulsating in fundamental mode while the P-\ks and the P-\wjk diagrams favour a relation to the first overtone pulsation on sequence C'. 
In all cases a small fraction of sequence (d) objects in the LMC, of the order of 10\%, seems to be shifted away from the bulk of the stars and possibly pulsate in a different pulsation mode. 
In the SMC the number of objects in this group is very small and it is difficult to reach conclusions.

\begin{figure}
\centering
\resizebox{\hsize}{!}{\includegraphics{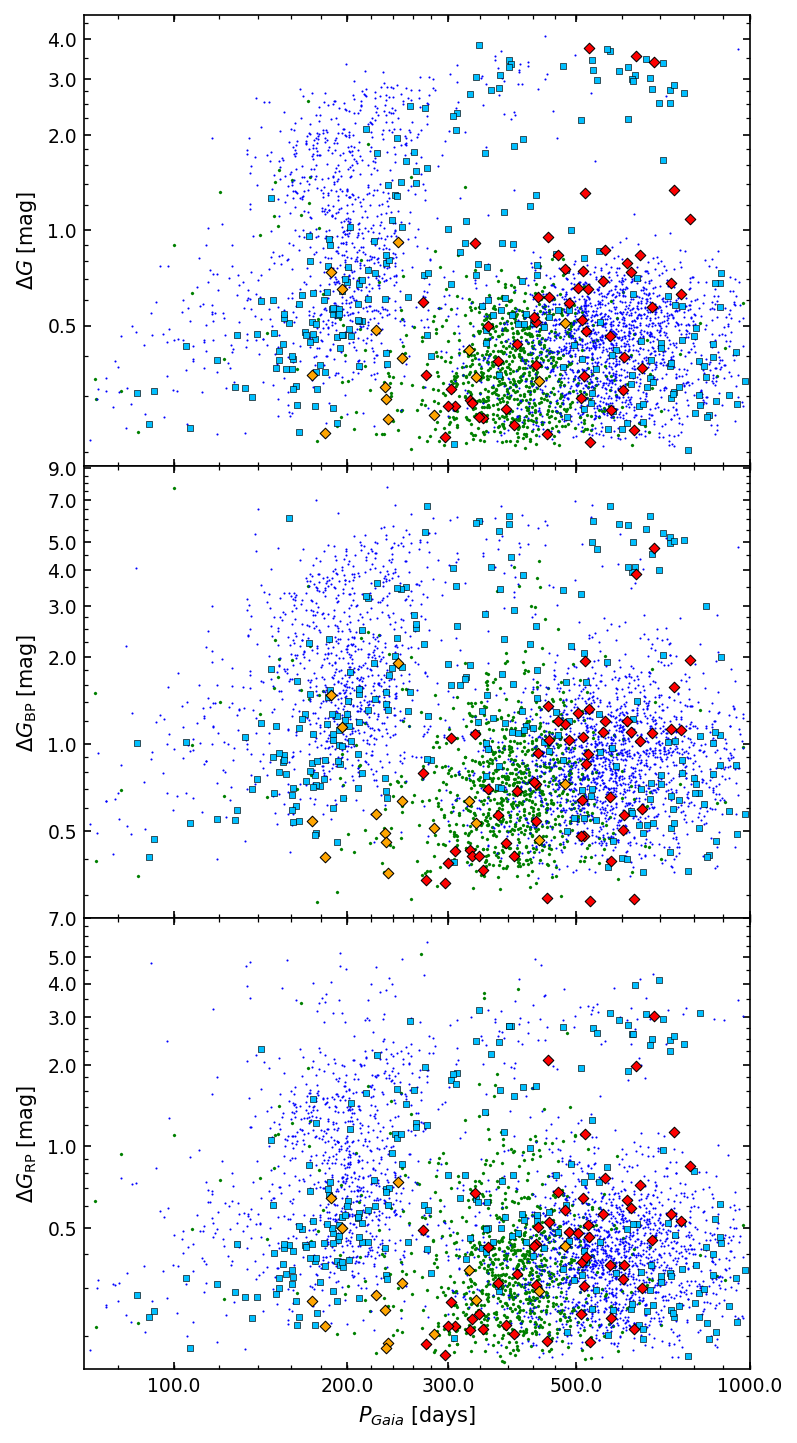}}
\caption{Period-amplitude diagrams of O-rich stars, colour coded by evolutionary group to which they belong. The green dots indicate faint AGBs and RGBs, the blue dots indicate low-mass AGBs, and the cyan squares indicate intermediate-mass AGBs. The massive AGBs and RSGs from branch (d) are shown as orange/red diamond symbols according to whether they belong to sequence C'/C in the P-\wrp diagram.
The three panels present the amplitudes in \gmag, \gbp, and \grp, respectively, taken from the magnitude ranges published in \Gaia DR2.}
\label{fig:AmplPerMassive}
\end{figure}

To investigate whether there are two groups of LPVs among the high-mass stars that can be distinguished by their pulsation mode, in Fig.~\ref{fig:AmplPerMassive} we compare the period with the amplitudes (taken from the magnitude ranges published in \Gaia DR2) in the \gmag, \gbp, and \grp bands for our complete sample.
In this diagram, massive AGB stars and RSGs are denoted according to the P-\wrp sequences to which they belong.
A difference in the pulsation amplitude may be a hint that different modes are excited.
For instance, there is an obvious gap in the amplitude distribution at approximately 2\,mag for P$>$300\,d.
The large amplitude stars form a band starting at a period of about 150\,d, and all these objects are located on sequence C and are fundamental mode pulsators.

It can be seen that all but three high-mass stars show very similar amplitudes of less than 2\,mag.
This result is independent of the photometric band used.
We conclude that the pulsation amplitudes do not support the presence of two groups with different pulsation modes among the high-mass LPVs.
The three objects with a significantly larger variability amplitude are all found close to sequence C in all luminosity indicators tested and their light curves show bumps in the rising parts of their light curves which are typical for massive Mira-like pulsators (see Fig.\,\ref{fig:rsg_example}).
We therefore conclude that these are massive AGB stars pulsating in fundamental mode.
In the P-\wrp diagram they are not distinguishable from the bulk of the stars on that sequence.

\citet{chatys_etal_2019} published a large collection of red supergiants from the galactic field and the LMC with their periods and visual amplitudes.
Like in our study, these authors found light amplitudes to be lower than 1 mag for the vast majority of objects. 
Only one star in their list, UZ CMa, which has been classified as a RV Tau variable, shows an amplitude of 2 mag. 
Even though the amplitudes reported in \citet{chatys_etal_2019} are in photometric bands different from those of \Gaia , we find our results to be in good agreement with that study.

While amplitudes alone do not allow us to draw a conclusion on the dominant pulsation mode, models suggest that higher mass tends to suppress overtone instability \citep[e.g.][their figure~22 and 23]{trabucchi_etal_2019}. 
A study of the velocity amplitude of typical representatives from this object class would add an important observational constrain to identify their pulsation modes.
If we assume the fundamental mode to be the dominant mode of pulsation in the red supergiants, models from \citet{wood_etal_1983} give typical masses between 15 and 30\,\Msun.

\begin{figure}
\centering
\resizebox{\hsize}{!}{\includegraphics{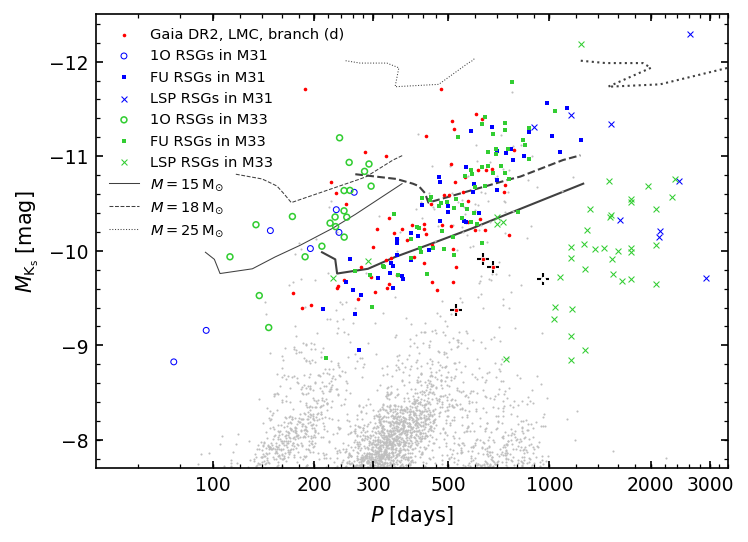}}
\caption{P-\ks diagram combining our \Gaia DR2 results for the LMC with observations of supergiants in M31 and M33 by \citet{ren_etal_2019}. Stars from \citet{ren_etal_2019} are indicated according to the pulsation mode given there, namely open symbols for first overtone pulsators and filled symbols for stars pulsating in fundamental mode. Long secondary periods are delineated with crosses. First overtone (thin lines) and fundamental mode (thick lines) periods from RSG pulsation models by \cite{guo_li_2002} are also shown, for a few different values of mass.}
\label{fig:M31M33PLD}
\end{figure}

In addition to the paper by \citet{chatys_etal_2019} mentioned above, \citet{ren_etal_2019} recently presented PLDs for supergiants in M31 and M33.
A comparison with pulsation models allowed the latter authors to attribute the supergiants to fundamental or first overtone pulsation.
In Fig.~\ref{fig:M31M33PLD} we plot the period and \ks values for supergiants derived by them together with the \Gaia DR2 results for the LMC.
It can be seen that the \Gaia group (d) stars fall on the relations defined by the M31 and M33 supergiants. 
A similar agreement was found with the PLDs presented by \citet{chatys_etal_2019} for galactic and LMC red supergiants.

From this plot, it also becomes clear that the LSPs of red supergiants occur on timescales of a few thousand days and are therefore out of reach for \Gaia DR2.
Using the conclusions of \citet{ren_etal_2019} on the pulsation mode of these objects by comparing their location in the P-\ks\  plane with pulsation models, we find the bulk of our group (d) stars to be fundamental mode pulsators. 
If we want to use the PLD to identify the pulsation mode of a star and if we attribute fundamental mode pulsation to sequence C, we must use the P-\wrp\  diagram because it is the only one with the massive stars falling on that sequence. 
For other indicators of luminosity such as \ks or \wjk, the shift likely resulting from the mass and temperature difference relative to low-mass stars is of a size that puts these stars onto sequence C'.
A parallel sequence shifted towards shorter periods, which is seen in our study (Fig.~\ref{fig:M31M33PLD}), in the paper by \citet{ren_etal_2019} and in the work by \citet{chatys_etal_2019}, is likely formed by first overtone pulsators among the massive AGB stars and the supergiants.

\section{Conclusions}
Gaia observations of the LMC and SMC
have allowed the study of the variability of red supergiants and massive AGB stars in
contrast to the extensive previous ground-based observations of OGLE and MACHO in which these
stars saturated the detectors used.
The \Gaia-2MASS diagram, constructed for the LMC, provides significant advantages in comparing the AGB and red supergiant populations of different stellar systems.
First of all, the quantity on the x-axis is independent of both distance and (global) reddening, and thus insensitive to the uncertainties in both quantities.
A comparison of the two plots in Fig.~\ref{fig:LMC_SMC_G2MD_apparent} suggests that the \Gaia-2MASS diagram is applicable to other galaxies as well.

Our diagram allows us to distinguish between stars of different mass and chemistry.
This paper focussed on the application of this distinction to the P-L relations of LPVs.
The appearance of the P-L relations depends on the spectral range of observations, so that different kinds of information can be derived from the PLD depending on the photometric pass-bands employed in its construction \citep[e.g.][their figure~1]{soszynski_etal_2007}. 
In this paper, we considered four such PLDs employing the $\ks$-band and $\wjk$ index from 2MASS, the $\wrp$ index from \Gaia DR2 photometry, and the bolometric magnitude \mbol, respectively.

We could, for the first time, clearly show the existence of an offset between low- and intermediate-mass, oxygen-rich stars for both fundamental and 1O pulsators and thus confirm predictions from pulsation theory.
The offset is, however, not visible when using the Wesenheit index \wrp. 
We explain this as the result of a temperature sensitivity inherent in the chosen combination of the \Gaia BP and RP filters.
The temperature effect is different for intermediate-mass and low-mass stars, and compensates the offset due to the mass effect.
As a consequence, sequence C' in the P-\wrp diagram is found to be the most narrow as compared to the other luminosity indicators investigated in this work.
Therefore, this relation may become the preferred approach to determine distances from the variability behaviour of AGB stars.
In addition, we possibly found a way to identify the most massive C-rich stars in our sample, using an analogy to the O-rich case.
However, this result requires validation by further tests and observational proof.

A detailed comparison with combined stellar evolution and pulsation models are presented, both for the O-rich and  C-rich cases. 
We showed that synthetic photometry from these models allows us to reproduce the observed behaviour of first overtone pulsators in all our four luminosity indicators very well.
As pointed out in other papers already, the linear pulsation models used in this work are not capable of reproducing the observations for fundamental mode pulsations. 

Our findings for massive AGB stars and red supergiants agree with former studies presented in the literature. 
Based on this, we agree with earlier claims that the majority of stars in this group pulsate in fundamental mode.
However, the amplitude of these fundamantal mode pulsating red supergiants never reaches the large values exhibited by intermediate- and low-mass stars during their final pulsating Mira stages.
In contrast to the P-\ks diagram and similar to the situation for intermediate-mass stars, the P-\wrp\  diagram based on \Gaia data shows this group of stars as a smooth extension of sequence C.
A few members of our group (d) seem to pulsate in first overtone mode according to their locations in the various PLDs.
For the supergiants, as for the other mass ranges, results for the two Magellanic Clouds are qualitatively similar.

\begin{acknowledgements}
The authors thank Franz Kerschbaum and Josef Hron for providing access to their photometric data archive used to derive bolometric correction relations. Furthermore, we thank Bernhard Aringer for support with stellar model spectra. MT, PM, and GP acknowledge the support from the ERC Consolidator Grant funding scheme ({\em project STARKEY}, G.A. n. 615604).\\
This work has made use of data from the European Space Agency (ESA) mission {\it Gaia} (\url{https://www.cosmos.esa.int/gaia}), processed by the {\it Gaia} Data Processing and Analysis Consortium (DPAC, \url{https://www.cosmos.esa.int/web/gaia/dpac/consortium}).
Funding for the DPAC has been provided by national institutions, in particular the institutions participating in the {\it Gaia} Multilateral Agreement.
This publication makes use of data products from the Two Micron All Sky Survey, which is a joint project of the University of Massachusetts and the Infrared Processing and Analysis Center/California Institute of Technology, funded by the National Aeronautics and Space Administration and the National Science Foundation.
This paper utilises public domain data obtained by the MACHO Project, jointly funded by the US Department of Energy through the University of California, Lawrence Livermore National Laboratory under contract No. W-7405-Eng-48, by the National Science Foundation through the Center for Particle Astrophysics of the University of California under cooperative agreement AST-8809616, and by the Mount Stromlo and Siding Spring Observatory, part of the Australian National University.
\end{acknowledgements}


\bibliographystyle{aa}
\bibliography{ReferencesPLD2}

\appendix
\section{Bolometric correction relation for \gmag}\label{sec:bc}
After having explored a BC on \Gaia photometry of LPVs in \citet{mowlavi_etal_2018_dr2lpv} using a small subset of the sample of reference objects from the galactic field presented in \citet{kerschbaum_etal_2010}, we extended the study for this paper to several hundred reference stars.
Spectral and variability type classifications were taken from the General Catalogue of Variable Stars (GCVS). 
A few stars without an existing spectral classification turned out to be primarily M-type stars according to their colours, so they were treated as M stars in our analysis.

Our sample for the BC analysis consisted of 466 M-type SRVs, 212 M-type Miras, 106 C-type SRVs, 33 C-type Miras, and 12 red supergiants, the selection of the latter being based on the spectral and variability classification in the GCVS. 
A wide range of colour and period characteristics were covered.
We first analysed each of the groups listed above separately, but since we did not find systematic differences between Miras and SRVs within each spectral class we decided to use only three correction formulas for M-type, C-type, and supergiants, respectively.
Bolometric correction relations were computed by comparing relative bolometric magnitudes from a fit to multi-band photometry from the optical to the far-infrared range \citep[see][for details]{kerschbaum_etal_2010} with \Gaia \gmag photometry as BC(G) $= \mbol - \gmag$.
The resulting BC values are thus negative and need to be added to \gmag. 
The \Gaia photometry is thereby a median of several measurements, while the ground-based multi-band photometry is typically a single-epoch measurement at a random phase.
This difference plays a negligable role for the semi-regular variables and supergiants, but may lead to a widening of the relations by the large amplitude Miras.
We further note that neither ground-based nor \Gaia photometry was corrected for interstellar reddening.
However, this simplification seems acceptable since we included only nearby LPVs in our study and since the fit of the bolometric flux was dominated by the near- and mid-infrared bands, which are much less affected by extinction.

For 678 M-type giants we derived a BC(\gmag) versus \gbp-\grp relation
\begin{align}\label{eq:BC_Mtype}
    {\rm BC}(\gmag) &= 0.790 - 0.953\,(\gbp-\grp) \\
    &+ 0.102\,(\gbp-\grp)^{2} - 0.014\,(\gbp-\grp)^{3} \,.\nonumber
\end{align}

\begin{figure}
\centering
\resizebox{\hsize}{!}{\includegraphics{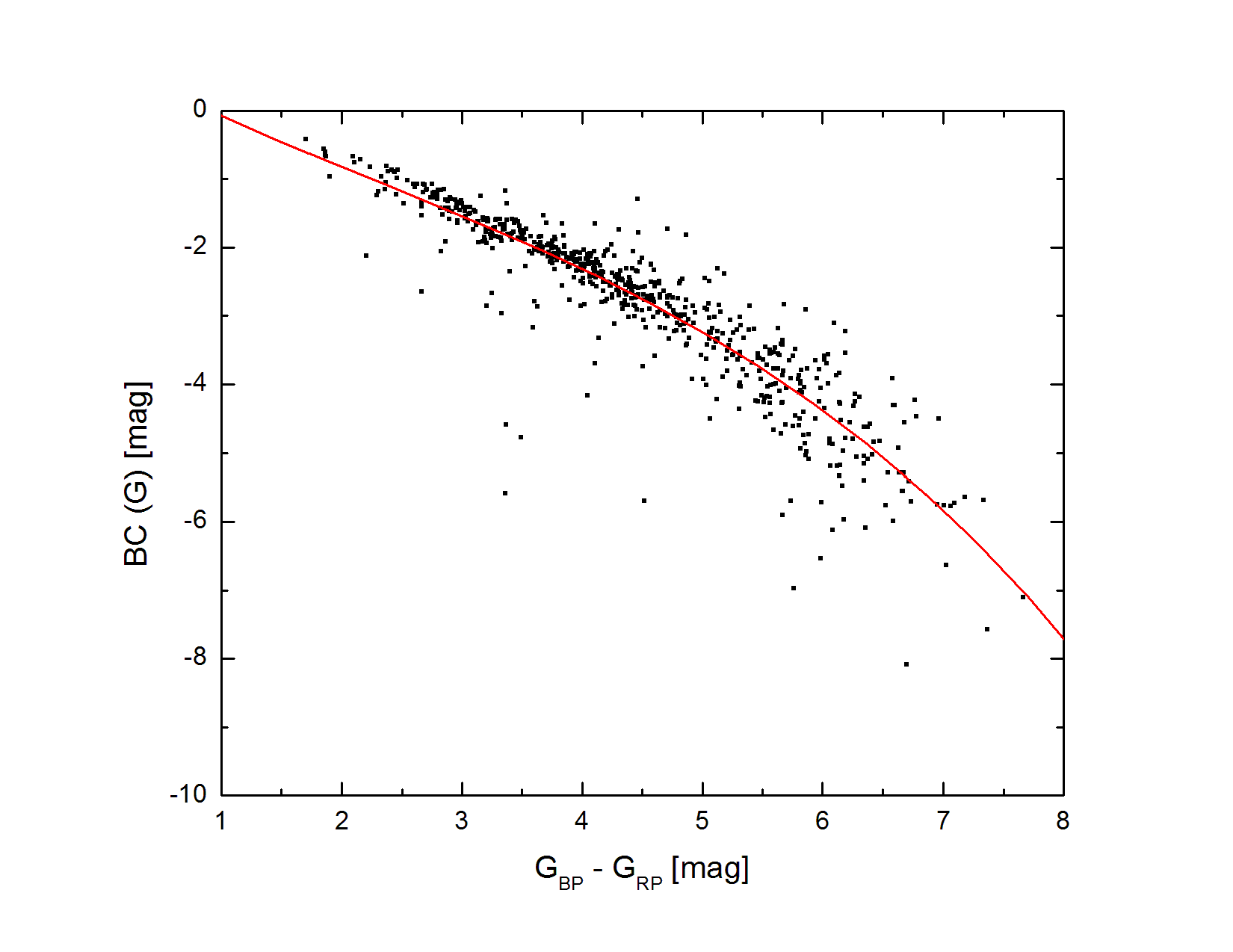}}
\caption{Bolometric corrections of \Gaia \gmag-band photometry for a sample of M-type red giants with \mbol values from \citet{kerschbaum_etal_2010}. The solid line shows the fitting curve presented in Eq.\,\ref{eq:BC_Mtype}. 
        }
\label{fig:BC_M}
\end{figure}

The reference objects used and the fitted relation can be seen in Fig.~\ref{fig:BC_M}. 
We decided for third degree polynomial fit as it allowed us to reproduce the observed bending for very red objects.
The sample used for the computation of the relation limits its applicability to a colour range $1.5<\gbp-\grp<7.5$.
This range covers the colour range of the LPV candidates in the \Gaia DR2 catalogue very well (see Fig.~13 of \citet{mowlavi_etal_2018_dr2lpv}).
For the very few targets outside of this \gbp-\grp range it seems feasible to extrapolate the derived relations.

The uncertainty of the BC is increasing with colour owing to the effects of variability and circumstellar reddening, which both increase for redder objects.
To quantify this uncertainty, we divided the \gbp-\grp colour range into bins of 0.1\,mag width and computed the standard deviation $\sigma$ of the BC value within each of these bins.
Then we made a linear fit of $\sigma$ as a function of \gbp-\grp.
The uncertainty of the BC for an M-type giant of a given colour can thus be computed as
\begin{align}\label{eq:BC_Mtype_sig}
    \sigma({\rm BC}) &= 0.134\,(\gbp-\grp) - 0.19 \,.
\end{align}
An interpolation of the uncertainty to colours outside the range of validity mentioned above is problematic owing to the linear relation used in this work.
To avoid getting negative error values for the bluest LPV candidates, we suggest setting the uncertainty to an arbitrary value of 0.01\,mag for stars with \gbp-\grp$<$1.5. 

In the same way we derived the following relation and uncertainty estimates for 139 C-type giants:
\begin{align}\label{eq:BC_Ctype}
    {\rm BC}(\gmag) &= -6.890 + 5.459\,(\gbp-\grp) \\
    &- 1.480\,(\gbp-\grp)^{2} + 0.057\,(\gbp-\grp)^{3}  \nonumber
\end{align}

\begin{align}\label{eq:BC_Ctype_sig}
    \sigma({\rm BC}) &= 0.634\,(\gbp-\grp) - 1.46. 
\end{align}

\begin{figure}
\centering
\resizebox{\hsize}{!}{\includegraphics{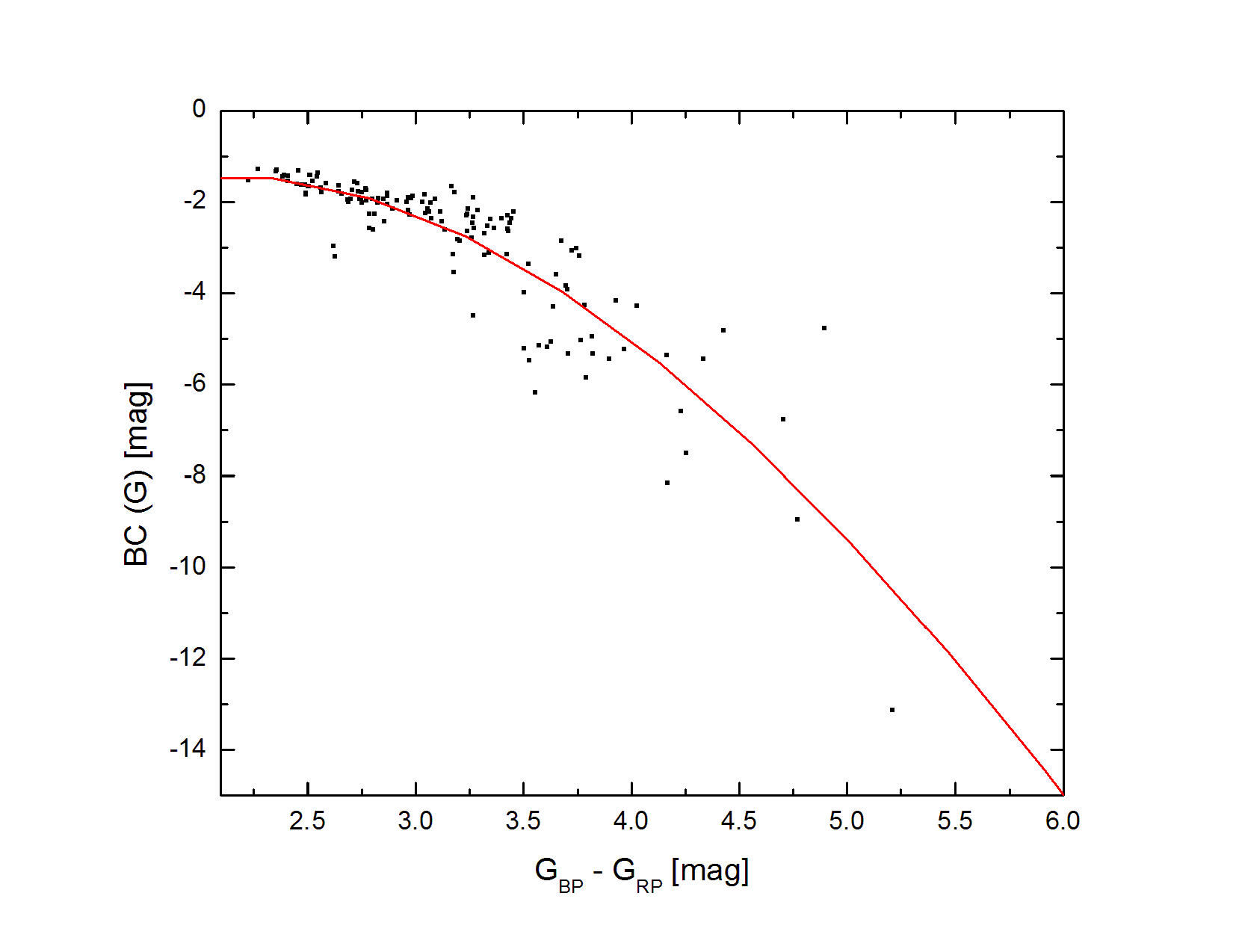}}
\caption{Bolometric corrections of \Gaia \gmag-band photometry for a sample of C-type red giants with \mbol values from \citet{kerschbaum_etal_2010}. The solid line shows the fitting curve presented in Eq.\,\ref{eq:BC_Ctype}. 
        }
\label{fig:BC_C}
\end{figure}

The data for the C-star sample and the fitted BC relation are presented in Fig.~\ref{fig:BC_C}.
Obviously, the relation is very different from that for the M giants. 
There is a very steep decline towards the red end. 
Even though we have to accept a somewhat higher uncertainty, the C-type Miras in our sample allow us to constrain the relation quite well. 
Differences to M-type stars occur also on the blue side of the relation, which makes a clear distinction of stars due to their chemistry indispensable for applying a proper BC.

\begin{figure}
\centering
\resizebox{\hsize}{!}{\includegraphics{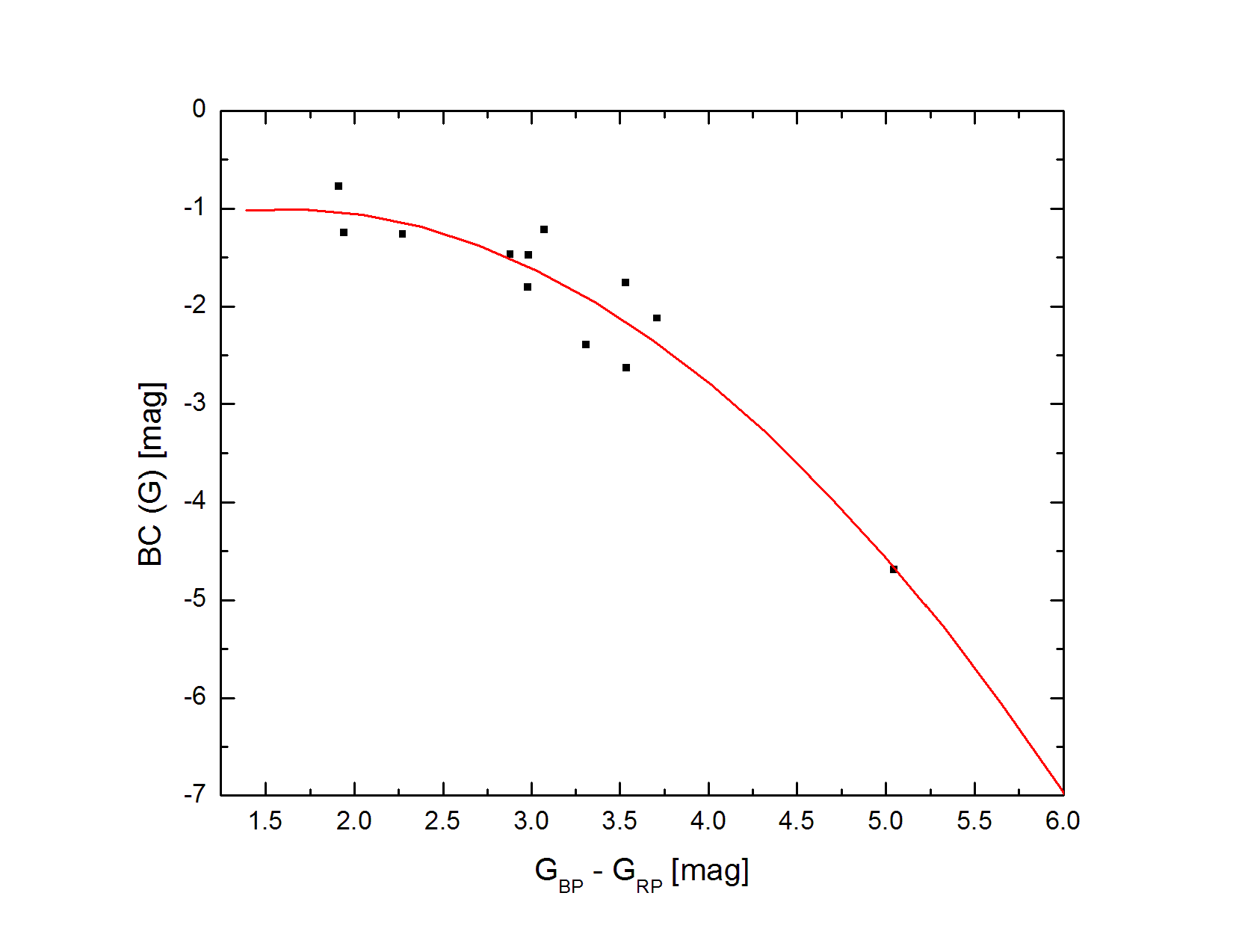}}
\caption{Bolometric corrections of \Gaia \gmag-band photometry for a sample of red supergiants with \mbol values from \citet{kerschbaum_etal_2010}. The solid line shows the fitting curve presented in Eq.\,\ref{eq:BC_RSG}. 
        }
\label{fig:BC_RSG}
\end{figure}

Finally, we constructed a BC relation for the red supergiants (Fig.~\ref{fig:BC_RSG}). 
While the number of reference objects available is comparably small, they seem to define a second order polynomial relation reasonably well, i.e.
\begin{align}\label{eq:BC_RSG}
    {\rm BC}(\gmag) &= -1.787 + 0.983\,(\gbp-\grp) \\
    &- 0.308\,(\gbp-\grp)^{2} \,.\nonumber
\end{align}

Because of the comparably small number of reference objects we decided not to derive an error function like in the previous two samples of O- and C-rich stars. 
Instead, we estimate a typical uncertainty of the BC for supergiants of 0.3 mag.

Tests for $G-\grp$ versus BC$(G)$ and $\gbp-G$ versus
BC$(G)$ relations were done, but $\gbp-\grp$ is preferred because the \Gaia BP and RP wavelength ranges have a very limited overlap, which is not the case between the \gmag and BP, or the \gmag and RP filters.
However, these alternative relations could be constructed for cases where measurements in one filter are missing.

\end{document}